\documentclass[preprint2,10pt,eqsecnum,flushrt]{aastex}
\usepackage{amsmath,amssymb,mathtools}
\usepackage{graphicx}
\usepackage{natbib}
\usepackage[usenames,dvipsnames]{color}
\usepackage{xspace}
\usepackage{soul}
\newcommand{\mycaption}[1]{\caption{\small #1}}
\newcommand{\ud}[1]{\,\mathrm{d}#1}
\newcommand{\br}[1]{\left({#1}\right)}
\renewcommand{\sq}[1]{\left[{#1}\right]}
\newcommand{\abs}[1]{\left|#1\right|}
\newcommand{\rsun}{\mathrm{r}_{\odot}}

\newcommand{\avg}[1]{\langle{#1}\rangle}
\newcommand{\avgg}[1]{\overline{#1}}

\newcommand{\myemph}[1]{\emph{#1}}

\newcommand{\mycitep}[1]{\citep{#1}}
\newcommand{\mycitet}[1]{\citet{#1}}

\newcommand{\disp}[1]{\langle{#1}\rangle}
\newcommand{\pdf}{\textsl{\small PDF}\xspace}
\newcommand{\rvd}{\textsl{\small RVD}\xspace}
\newcommand{\lsr}{\textsl{\small LSR}\xspace}
\newcommand{\msun}{\mathrm{M}_{\odot}}

\newcommand{\kpc}{\,\mathrm{kpc}}

\renewcommand{\d}[2]{{#1}'{(#2)}}
\newcommand{\pdd}[2]{\frac{\partial L}{\partial\d{#1}{#2}}}
\newcommand{\pb}[2]{\left\{#1,#2\right\}}
\newcommand{\pd}[2]{\partial_{#2}{#1}}
\newcommand{\secref}[1]{Sect.\,\ref{#1}}

\newcommand{\figref}[1]{Fig.\,\ref{#1}}
\renewcommand{\eqref}[1]{Eq.\,\ref{#1}}
\newcommand{\donotremove}[1]{}
\newcommand{\oka}{$\mathrm{O}_{_{\mathrm{VII}}}\mathrm{K}_{\alpha}$\xspace}
\newcommand{\mref}{M_{\mathrm{ref}}}
\newcommand{\spc}{@{\hspace{3pt}}}

\begin{document}

\title{Motion of halo compact objects in the gravitational potential  of a low-mass model of the Galaxy}

\shorttitle{Motion of compact objects in a low-mass Galaxy model}
\shortauthors{S. Sikora, {\L}. Bratek, J. Ja{\l}ocha, M.
Kutschera}

\author{Szymon Sikora}
\affil{Astronomical
Observatory, Jagellonian University, Orla 171, PL-30244
Krak{\'o}w, Poland}

\author{{\L}ukasz Bratek}
\affil{Institute of Nuclear Physics, Polish Academy of Sciences,
Radzikowskego 152, PL-31342 Krak\'{o}w, Poland}
\email{Lukasz.Bratek@ifj.edu.pl}

\author{Joanna Ja{\l}ocha}
\affil{Institute of Nuclear Physics, Polish Academy of Sciences,
Radzikowskego 152, PL-31342 Krak\'{o}w, Poland}

\author{Marek Kutschera}
\affil{Institute of
Physics, Jagellonian University,  Reymonta 4, PL-30059 Krak{\'o}w,
Poland}

\begin{abstract}
Recently, we determined a lower bound for the Milky Way mass in a point mass approximation. This result was obtained 
for most general spherically symmetric phase-space distribution functions consistent with a measured radial velocity 
dispersion. As a stability test of these predictions against a perturbation of the point mass potential, in this paper 
we make use of a representative of these functions to set the initial conditions for a simulation in a more realistic potential of similar mass and accounting for other observations.
The predicted radial 
velocity dispersion profile evolves to forms still consistent with the measured profile, proving structural stability of the 
point mass approximation and the reliability of  the resulting mass  estimate of $\sim2.1\!\times\!10^{11}\msun$ within $150
\kpc$. We also find an interesting coincidence with the recent estimates based on the kinematics of the extended Orphan Stream. 
As a byproduct, we obtain the equations of motion in axial symmetry from a nonstandard  Hamiltonian, and derive a formula in the spherical symmetry relating the radial velocity dispersion profile 
to a directly measured kinematical observable.
\end{abstract}


\maketitle

\section{\label{sec:intro}Introduction}
Compact objects orbiting the Milky Way can be used to infer the gravitational field at large Galactocentric radii. 
Jeans modeling links the field
with the available kinematical data, 
under the important assumption that these 
objects can be described by a collision-less system of test bodies in a steady-state equilibrium \mycitep{1915MNRAS..76...70J}. 
Primary in this approach is a \textsl{\small P}\textit{hase-space} \textsl{\small D}{istribution} \textsl{\small F}\!\textit{unction} (\myemph{\pdf}), while 
the physical observables (e.g., the number density, the 
velocity dispersion ellipsoid, etc.) are \myemph{secondary quantities} being \pdf-dependent functionals on the phase space. 

Galaxy mass can be estimated based on the \textsl{\small R}{\textit{adial}} \textsl{\small V}\!\textit{elocity} \textsl{\small D}\textit{ispersion} (\myemph{\rvd}) data. 
In the literature, one can find mass values of $4.2_{-0.4}^{+0.4}\!\times\!10^{11}\msun$ \mycitep{2012MNRAS.424L..44D}, 
$4.9_{-1.1}^{+1.1}\!\times\!10^{11}\msun$ \mycitep{1996ApJ...457..228K}, $5.4_{-3.6}^{+0.2}\!\times\!10^{11}\msun$ \mycitep{1999MNRAS.310..645W} and  $5.4_{-0.4}^{+0.1}\!\times\!10^{11}\msun$ \mycitep{2003A&A...397..899S}, all within $50\kpc$,  $4.0_{-0.7}^{+0.7}\times10^{11}\msun$ within $60\kpc$ \citep{2008ApJ...684.1143X}, or
$(5.8\!-\!6.0)\!\times\!10^{11}\msun$ enclosed within $100\kpc$  \mycitep{2002ApJ...573..597K}, whereas, depending on the model assumptions, 
the virial mass is $(8\!-\!10\!-\!12)\!\times\!10^{11}\msun$ \mycitep{2005MNRAS.364..433B,2008ApJ...684.1143X,2014ApJ...794...59K} or even
$(18-25)\times10^{11}$ \citep{2003A&A...397..899S}. 
It is noteworthy that the  kinematics of extended Orphan Stream indicates the mass of $\sim2.7\!\times\!10^{11}\msun$ within $60\kpc$ (with disk+bulge mass of $1.3\times\!10^{11}\!\msun$) \citep{2010ApJ...711...32N,2013ApJ...776...26S},   significantly less than suggested by the above estimates within $50\kpc$. This, points to some model-dependent effects.

In this context it is natural to ask what is the lower bound for the Galaxy mass indicated by the kinematics of the outermost tracers, 
paying more attention on the phase space model rather than the particular mass model. 

With the simplest working hypothesis of absence of the extended halo, the Galactic gravitational field at large distances, to a fair degree of approximation, would be that of a point mass.
In this approximation, decisive for the field asymptotics of any compact mass distribution is a single total mass parameter.  \mycitet{1981ApJ...244..805B} proposed in this approximation for the neighboring galaxies  a mass estimator $\frac{C}{G}\left<v_z^2\,R \right>$, with the averaging performed over distant tracers at various projected radii $R$, where $C$ is a constant. 
More recently, \mycitet{2010MNRAS.406..264W} considered a spherical symmetric counterpart of this estimator, $\frac{C}{G}\avg{v_r^2\,r^{\gamma}}$, 
with an arbitrary power of the radial distance $r$. 

Mass estimators based on Jeans theory, irrespectively of the addopted mass model, are related to a \pdf restricted in a particular way by some indirect constraints appearing because of the assumptions made
about the secondary quantities. 
The restrictions usually concern {the flattening of the velocity dispersion ellipsoid} $\beta$.
This quantity is poorly  known for peripheral 
tracers. Introducing a variable $\beta$ leads to difficulties in solving the Jeans problem. 
To overcome this, $\beta$ is often assumed to be a position independent parameter.
On the other hand, this is too much constraining an assumption, since any limitation on  $\beta$ indirectly imposes restrictions on the function space admissible for \pdf's. We conjectured \mycitep{2014A&A...562A.134B} that the lower bound for the total mass may increase in response to these constraints, while there is no definite upper bound. In consequence of this the mass is likely to be  overestimated. 
 
Recently, there is a growing interest in methods of determining the general assumption-free \pdf{s} from the kinematical data. 
\mycitet{2014MNRAS.437.2230M} proposed a framework in which the gravitational potential is inferred from a discrete realization of the unknown distribution function by using snapshots of stellar kinematics.
Our previous article \mycitep{2014A&A...562A.134B} is placed within this field of interest. Therein, we proposed  a method of determining \pdf's from a given spherically symmetric \rvd profile, without imposing any constraints on the secondary quantities. Even in the simplest case of a point mass approximation, our method allowed 
to faithfully reconstruct the shape of the \rvd profile, including its low-size and variable features. 
By considering  various \pdf's giving rise to \rvd's overlapping with that observed at larger radii, we showed 
that there is no upper bound for the total mass, while
there is a sharp lower bound, slightly below $2.0\!\times\!10^{11}\msun$. 
For lower masses, 
no \pdf could 
be found to account for 
the measured \rvd within the acceptable limits. 

\subsection{The aim of the present work}

The lower bound referred to above may also suggest a Galaxy mass lower 
than 
given in the literature. A natural question arises whether 
it is physically reasonable and could appear in other models? 
This cannot be excluded. 
A mass 
$(2.4\!-\!2.6)\!\times\!10^{11}$ would be consistent with past results -- 
in a three-component
mass model \citep{1992AJ....103.1552M}, with the best estimate in the point mass field 
\citep{1987ApJ...320..493L} --
and most remarkably, with the recent value inferred from the kinematics of the Orphan stream \citep{2010ApJ...711...32N,2013ApJ...776...26S}.

Interestingly, the lower bound coincides with the sum of the dynamical mass  $\approx1.5\!\times\!10^{11}\msun$ inferred from the rotation curve in disk model \mycitep{2014A&A...566A..87J, 2012A&A...546A.126S} and the mass $(1.2\!-\!6.1)\!\times\!10^{10}\msun$  of the hot gaseous halo surrounding Milky Way 
\mycitep{2012ApJ...756L...8G}. 
The gravitational potential of these components can be interpreted as a perturbation of a point mass potential. But for complicated potentials, the distribution integral on the phase space cannot be explicitly constructed, because the first integrals characterizing admissible orbits are not known in an explicit form. To overcome this difficulty, a numerical simulation can be performed. 

In this paper we present an example of such a simulation in which the test bodies represent the compact objects orbiting the Galaxy. The initial conditions for the simulation in the perturbed field should be chosen close to a known stationary solution of Jeans problem in the non-perturbed field. For this purpose we make use of a \pdf to be found similarly as in \mycitep{2014A&A...562A.134B}. But it is not obvious if this initial \pdf and the resulting \rvd profile would be stable against such a perturbation. Running a simulation with an initial \pdf consistent with the observed high \rvd could lead to an \rvd profile   
with quickly and steadily decreasing value. If this happened, this would mean that the initial approximation was far from a stationary solution in the new potential. The main goal of the present paper is to investigate 
this stability issue and thus also the reliability of the point mass approximation, that is, 
we test whether predictions for the \rvd and the mass in the new potential are comparable to those made in the point mass approximation, that is, we test {\it structural stability} of these predictions.

\medskip

The structure of this paper is the following. 
In \secref{sec:model} we recall the main ideas behind the \textit{Keplerian ensemble method}  of obtaining a \pdf. 
In \secref{sec:simul},  we use a representative of possible \pdf's to set the initial conditions for our n-body simulation in the modified potential, next we test stability of the resulting \rvd profile. 
Then conclusions follow.
 
\section{\label{sec:model}First approximation of \pdf from the Keplerian Ensemble Method}

{\noindent \bf Mathematical preliminaries.}\footnote{This paragraph summarizes the mathematical basis of our method discussed in a more detail in \mycitep{2014A&A...562A.134B}.}
An elliptic orbit of a test body bound  in the field of a point mass $M$ is fully characterized by 5 integrals of motion: the Euler angles $(\Phi,\Theta,\Psi)$ determining the orbit orientation, the eccentricity $e$ describing the orbit flattening, and the dimensionless energy parameter $\epsilon=-\frac{RE}{GM}$ describing the size of the large semi-axis ($R$ is an arbitrary unit of length while $E$ is the energy per unit mass). A spherically symmetric collection of confocal ellipses we call  \textit{Keplerian ensemble}. 
From \mycitet{1915MNRAS..76...70J} theorem, for this ensemble in a steady state equilibrium it suffices to consider  \pdf's being functions of  $e$ and $\epsilon$ only.
Accordingly, instead of  $r$, $\theta$, $\phi$, $v_r$, $v_\theta$, $v_\phi$ we use phase coordinates $u$, $\theta$, $\phi$, $e$, $\epsilon$, $\psi$ defined by: 
\begin{equation}\label{transformacja}
\begin{aligned}
&{\textstyle r=R\,u,\qquad 
v_r^2=\frac{GM}{R}\left(\frac{2}{u}-\frac{1-e^2}{2\,\epsilon \,u^2}-2\,\epsilon \right),}\\
&{\textstyle (v_\theta,v_\phi)=\frac{1}{u}\,\sqrt{\frac{GM}{R}}\left(\frac{1-e^2}{2\,\epsilon}\right)^{1/2}\,(\sin\psi,\cos\psi)}.
\end{aligned}
\end{equation}
For physical reasons we assume all orbits to be confined entirely 
within a region $u\in(u_a,u_b)$ bounded by two spheres of radii $u_a$ and $u_b$.
This way all  orbits with too low pericentra (i.e., violating the point mass approximation) or with too high apocentra (e.g. beyond Local Group members)
are excluded. Consequently, the space of parameters $(e,\epsilon)$ gets 
restricted to a domain $S$: 
$\frac{1+e}{2u_b}<\epsilon<\frac{1-e}{2u_a}$ and $0\leq e<\frac{u_b-u_a}{u_b+u_a}<1$. On integrating out the angles $\theta,\phi,\psi$, the  principal  integral $\int f(\vec{r},\vec{v})\ud{}^3\vec{r}\ud{}^3\vec{v}$ reduces 
(to within an unimportant constant factor) to
\begin{equation}\label{funkcja_nu}
\begin{aligned}
 \int\limits_{u_a}^{u_b}\!\nu_u[f]\ud{u},
\qquad 
\nu_u[f]\equiv\!\!\!\int\limits_{S(u)}\!\!\frac{e\,\mathrm{d} e\,\mathrm{d}\epsilon\,f(e,\epsilon)}{\sqrt{\epsilon\,\left(\epsilon-\frac{1-e}{2u}\right)\left(\frac{1+e}{2u}-\epsilon\right)}} \,.
\end{aligned}
\end{equation}
The integration domain {$S(u)\subset S$} is a $u$-dependent quadrilateral region
{\scriptsize$\epsilon\!\in\!\br{\mathrm{Max}\!\br{\frac{1-e}{2u},\frac{1+e}{2u_b}}\!,\mathrm{Min}\!\br{\frac{1+e}{2u},\frac{1-e}{2u_a}}\!}$}, {\scriptsize$e\!\in\!\br{0,\frac{u_b-u_a}{u_a+u_b}}$}, each point of which corresponds to a spherically symmetric pencil of confocal elliptical orbits
intersecting a sphere of a certain radius $u$. 
The functional $\nu_u[f]$ has the interpretation of the probability density for the variable $r/R$ to fall  within the spherical shell $u<r/R<u+\ud{u}$. 

Given a \pdf $f(e,\epsilon)$, the expectation value $\avg{g}_r$ for an observable $g=g(e,\epsilon,u)$ inside that shell  equals
\begin{equation}\label{srednia}
\avg{g}_r=\frac{\nu_u[fg]}{\nu_u[f]}\,.
\end{equation}
In particular, given $M$ and $f(e,\epsilon,u)$, the model \rvd profile $\avg{r\,v_r^2}/G$ is obtained with
$g(e,\epsilon,u)\equiv\frac{r\,v_r^2}{G\,M}=2-\frac{1-e^2}{2\,\epsilon \,u}-2\,\epsilon\,u\,$ substituted for $g$ in \eqref{srednia}.\footnote{
In place of $(e,\epsilon,u)$ it is more convenient the use of
 coordinates $(\alpha,\beta,u)$, such that 
$\textstyle e\!=\!\frac{\beta-\alpha}{\beta+\alpha}$,
$\epsilon\!=\!\frac{1}{u}\,\frac{\alpha\beta}{\alpha+\beta}$.
A motivation behind this mapping and its explicit construction is given in \mycitep{2014A&A...562A.134B}.} 

But we are concerned with the inverse problem: given an \rvd profile matching the observations, we want to derive a distribution function $f$ the \rvd profile would follow from. This problem can be solved as follows.
First, we consider an auxiliary function $h(e,\epsilon)$ such that $f\!=\!h^2$ (then $f\!\geq\!0$, as required for a probability density) and make a series expansion in  polynomials $\mathcal{Q}_k$ orthogonal on $S$:
\begin{equation}\label{eq:series}{\textstyle h(e,\epsilon)\approx\sum_{k=1}^{D_d}h_k\,\mathcal{Q}_k(e,\epsilon)}. \end{equation}
The $\mathcal{Q}_k$'s are constructed with the help of a Gramm-Schmidt orthogonalization method on $S$. Next, given a mass parameter $M$, we find an optimum  sequence of expansion coefficients $h_k$ by minimizing a discrepancy measure between {\it a)} the \rvd profile from measurements,  $\avgg{p}_r\equiv{\avgg{r v_r^2}}/G$, where the averaging is taken over all halo compact objects within a spherical shell of some width and a given radius $r$,  and {\it b)} the model $\avg{g}_r$ profile calculated from \eqref{srednia} with the help of the function $h$ corresponding to the optimum $h_k$'s. With these $h_k$'s,  the discrepancy measure can be reduced further by replacing $M$ with a better fit value, e.g., $M\!\to\! M_{bf}\!=\!\frac{\sum_r\avg{g}_r \avgg{p}_r}{\sum_r\avg{g}_r^2}$ if the $\sum_r\!\br{M\avg{g}_r\!-\!\avgg{p}_r}^2$ norm is used.  

This way, a \pdf $f(e,\epsilon)$ consistent with the \rvd measurements can be reconstructed, provided $M$ is large enough. 
For $M$ above a limiting value $M_{\mathrm{cut}}$ there is always  a \pdf for which the \rvd profile is perfectly accounted for, while  below this limit no satisfactory fit can be found. For $M\!>\!M_{\mathrm{cut}}$, increasing the number $D_d$ of the basis polynomials $\mathcal{Q}_k$, efficiently decreases the fit residuals, but for $D_d$ high enough the residuals appear to tend to some small nonzero limit. 
For $M\!<\!M_{\mathrm{cut}}$ the fit residuals remain very large, regardless of $D_d$, and rapidly increase with decreasing $M$. This shows  that $M_{\mathrm{cut}}$ is the lower bound for the mass in the point mass approximation. 

\subsection{\rvd profile from measurements}
Without transverse velocity components, the radial motions of kinematic tracers cannot be unambiguously transformed from the \lsr frame  to the Galactocentric frame. 
However, for a spherically symmetric distribution of tracers
one can try to assume a $\beta(r)$ profile or find a self-consistent one by iterations.
The particular model of $\beta(r)$ affects the \rvd significantly only inside a spherical region of several $\rsun$ in diameter.
But we must bear in mind a twofold influence of the particular model of $\beta$ on the total mass determination:  both $\beta(r)$ itself and the so obtained $\beta$-dependent \rvd profile enter the spherical Jeans equations.  

\subsubsection{A formula relating the \lsr radial motion measurements to the Galacto-centric \rvd}
Here, we consider a spherically symmetric ensemble of test bodies described by some \pdf and the resulting $\beta(r)$,  then also  $\avg{v_{\phi}^2(r)}=\avg{v_{\theta}^2(r)}$ (averaging over spherical shells). For a test body with a velocity vector $\vec{v}$ in the Galacto-centric coordinate frame, the radial and tangential components of $\vec{v}$ are
 $v_{r}=\vec{v}\circ\vec{e}_r$, $v_{\theta}=\vec{v}\circ\vec{e}_\theta$, $v_\phi=\vec{v}\circ\vec{e}_\phi$, with
$\vec{e}_r=\frac{\vec{r}}{|\vec{r}|},\vec{e}_\theta,\vec{e}_\phi$ forming an orthonormal basis tangent to the lines of constant spherical coordinates $r,\theta,\phi$. Although $\vec{v}$ can be determined for closer objects, only its projection $\tilde{v}_{r}=\vec{v}\circ\vec{e}_{\varrho}$ onto 
the line of sight  
determined by the unit vector  $\vec{e}_{\varrho}=\frac{\vec{r}-\vec{r}_{\odot}}{|\vec{r}-\vec{r}_{\odot}|}$ can be measured for all objects. 
This is the only kinematical information available at large distances, suitable for constraining the total Galactic mass.
It is connected with direct measurements of the \lsr relative velocity $v_{\varrho}$ along the direction $\vec{e}_{\varrho}$ through the relation 
$\tilde{v}_r=v_{\varrho}+\vec{v}_{\odot}\circ\vec{e}_{\varrho}$.  

Assuming a $\beta(r)$, we can relate  $\avg{\tilde{v}_r^2(r)}$ to $\avg{{v}_r^2(r)}$ through the following identity true both for $r<\rsun$ and $r>\rsun$:
\begin{eqnarray}\label{eq:mytransform}
\begin{aligned}
\avg{\tilde{v}_r^2(r)}&=\avg{v_r^2(r)}\br{1-\frac{\beta(r)}{4}H\br{{r}/{\rsun}}}\\
H\br{x}&= {
1+x^{-2}-\frac{\br{x^2-1}^2}{2\,x^3}\ln\abs{\frac{x+1}{x-1}}
}.\end{aligned}\end{eqnarray}
For the isotropic dispersion, $\beta(r)\!=\!0$,
we can formally make the (incorrect) identification $\tilde{v}_r\!=\!{v}_r$ globally, 
without making any error in equating the resulting dispersions $\avg{\tilde{v}_r^2}\!=\!\avg{v_r^2}$. 
For large $r$, $H\!\sim\!1-\beta(r)\frac{2\rsun^2}{3r^2}\!\to\!1$ if $\beta$ is asymptotically bound (which may not hold for nearly circular orbits). 
\mycitet{2005MNRAS.364..433B} derived a similarly looking result only for $r>\rsun$, concluding  -- counter to \eqref{eq:mytransform} and the geometric intuition -- that $\avg{{v}_r^2(r)}$ and $\avg{\tilde{v}_r^2(r)}$ coincided for purely radial anisotropic ellipsoid ($\beta\!=\!1$), not for purely isotropic ones ($\beta\!=\!0$). To dispel doubts as to which conclusion is correct, we present our derivation of  \eqref{eq:mytransform} in Appendix \ref{app1}.

While determining a \pdf $f\!=\!f(\vec{r},\vec{v})$ from the $\avg{\tilde{v}_r^2(r)}$ observable, a self-consistent $\beta(r)$ may be looked for by iterations.
The first recursion step makes the 
assumption $\avg{{v}_r^2(r)}\!=\!\avg{\tilde{v}_r^2(r)}$ as if $\beta(r)\!=\!0$, and a first approximation to $f$ is obtained, from which a $\beta(r)$  prediction  for the next iteration step is calculated. Substituted in \eqref{eq:mytransform}, the $\beta(r)$ 
gives rise to a new  $\avg{{v}_r^2(r)}$. The process is repeated untill a stable $\beta(r)$ is reached. However, 
the distinction between $\avg{{v}_r^2(r)}$ and $\avg{\tilde{v}_r^2(r)}$ is practically unimportant, unless the lower radii region 
is considered. In preparing the \rvd profile below, we may neglect this distinction.

\subsubsection{\label{sec:measurements} Measurements data}

\begin{figure}[h]
\centering
\includegraphics[trim = 16pt 1pt 12pt 0pt,
clip,
width=0.46\textwidth]{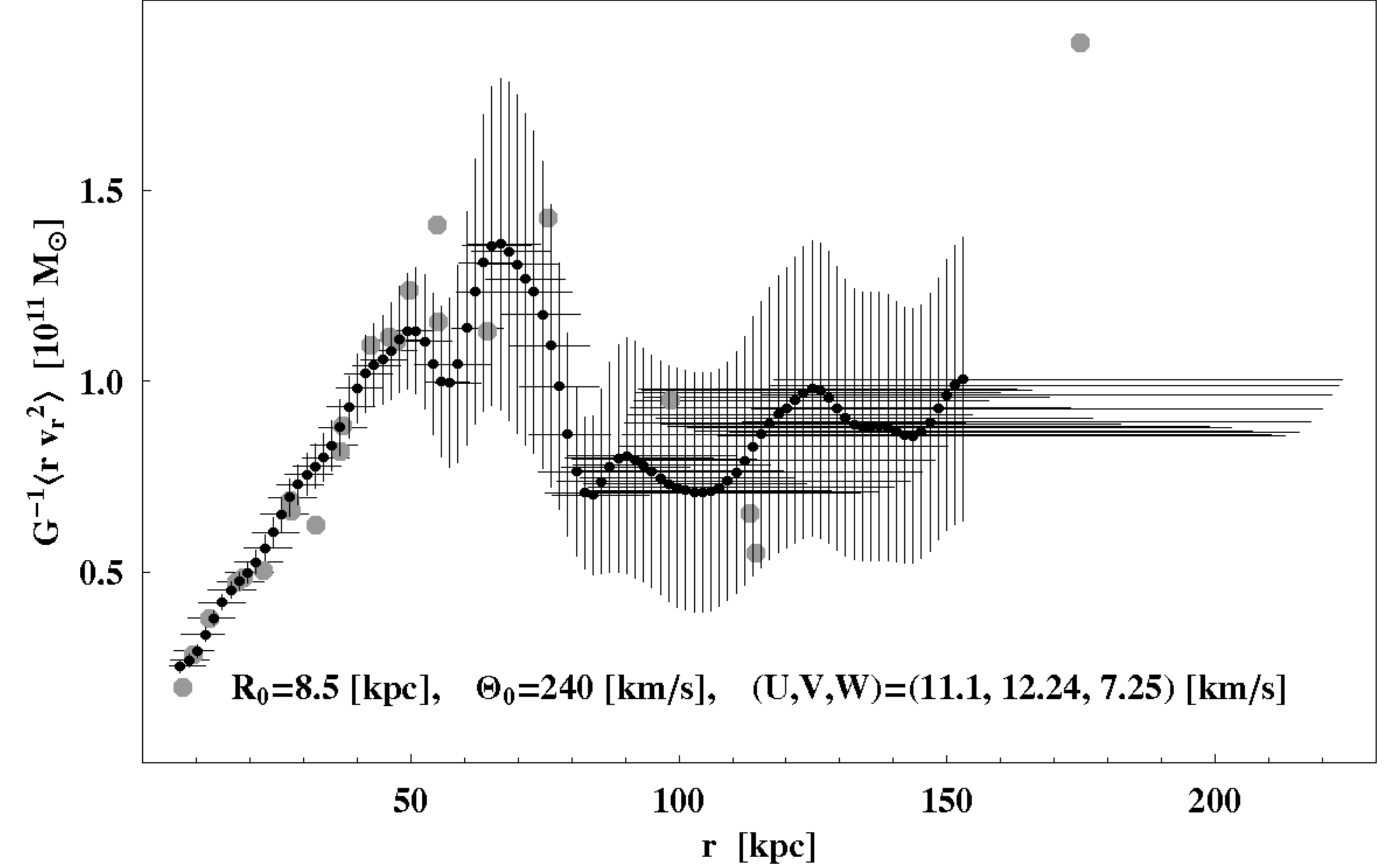}
\mycaption{\label{fig:RVDprofiles} \rvd profile $G^{-1}\disp{rv_r^2}$  for tracers with
$\frac{1}{2G}rv_r^2\lesssim3.5\!\times\!10^{11}\msun$.
The horizontal bars represent the
effective radial bin size of the moving average. The vertical bars
indicate the spread in the profile due to the inclusion/exclusion of
random subsets of tracers. A detailed description of how this
profile was obtained is given in \mycitep{2014A&A...562A.134B}.
As reference values, we show $G^{-1}r\disp{v_r^2}$ calculated based
on the \rvd points in \mycitep{2008ApJ...684.1143X} and
\mycitep{2012MNRAS.425.2840D} \textit{[large gray circles]}
}
\end{figure}

In our previous work \mycitep{2014A&A...562A.134B} we determined a \rvd profile \figref{fig:RVDprofiles} which we now assume as the basis for generating the initial conditions for the simulation in \secref{subsec:simul}, and use it as a reference profile for comparison with the simulation results in \secref{sec:results}.
We obtained this profile with the use of the following position-velocity data: the halo giant stars \mycitep{2001ApJ...555L..37D,2009ApJ...698..567S} from the Spaghetti Project Survey \mycitep{2000AJ....119.2254M}; the blue horizontal branch stars \mycitep{2004MNRAS.352..285C} from the United Kingdom Schmidt Telescope observations and SDSS; the field horizontal branch and A-type stars \mycitep{1999AJ....117.2329W} from the \mycitet{1992AJ....103.1987B} survey; the globular clusters \mycitep{1996AJ....112.1487H} and the dwarf galaxies \mycitep{1998ARA&A..36..435M}. The data were recalculated to epoch J2000 when necessary. In addition, we included the ultra-faint dwarf galaxies such as Ursa Major I and II, Coma Berenices, Canes Venatici I and II, Hercules \mycitep{2007ApJ...670..313S}, Bootes I, Willman 1 \mycitep{2007MNRAS.380..281M}, Bootes II \mycitep{2009ApJ...690..453K}, Leo V \mycitep{2008ApJ...686L..83B}, 
Segue I \mycitep{2009ApJ...692.1464G}, and Segue II \mycitep{2009MNRAS.397.1748B}. 
To eliminate a possible decrease in 
the \rvd at lower radii due to 
circular orbits in the disk, we excluded tracers in a neighborhood 
$(R/20)^2+(Z/4)^2<1$ (in units of $\kpc$)
of the mid-plane. 
We also did not take into account:  {\it a)} a distant Leo T located at $r > 400\,\mathrm{kpc}$,  
{\it b)}  Leo I, rejected for reasons largely discussed in \mycitep{2014A&A...562A.134B}, {\it c)} a single star for which $r v_r^2/(2G)>5.6\!\times\!10^{11}\msun$, and {\it d)}
$4$ additional objects
for which $r\,v_r^2/(2G)\gtrsim3.5\!\times\!10^{11}\msun$ (these are: {\small 88-TARG37, Hercules, J234809.03-010737.6} and {\small J124721.34+384157.9}).  As shown  with the help of a simple asymptotic estimator \mycitep{2014A&A...562A.134B}, had we not excluded {\it d)} the total expected mass would have been increased by only a factor of $\approx1.16$. 

\section{\label{sec:simul}Simulation of \rvd in a background field}

\begin{figure}[h]
\centering
\includegraphics[trim = 5pt 3pt 6pt 2pt,
clip,
width=0.46\textwidth]{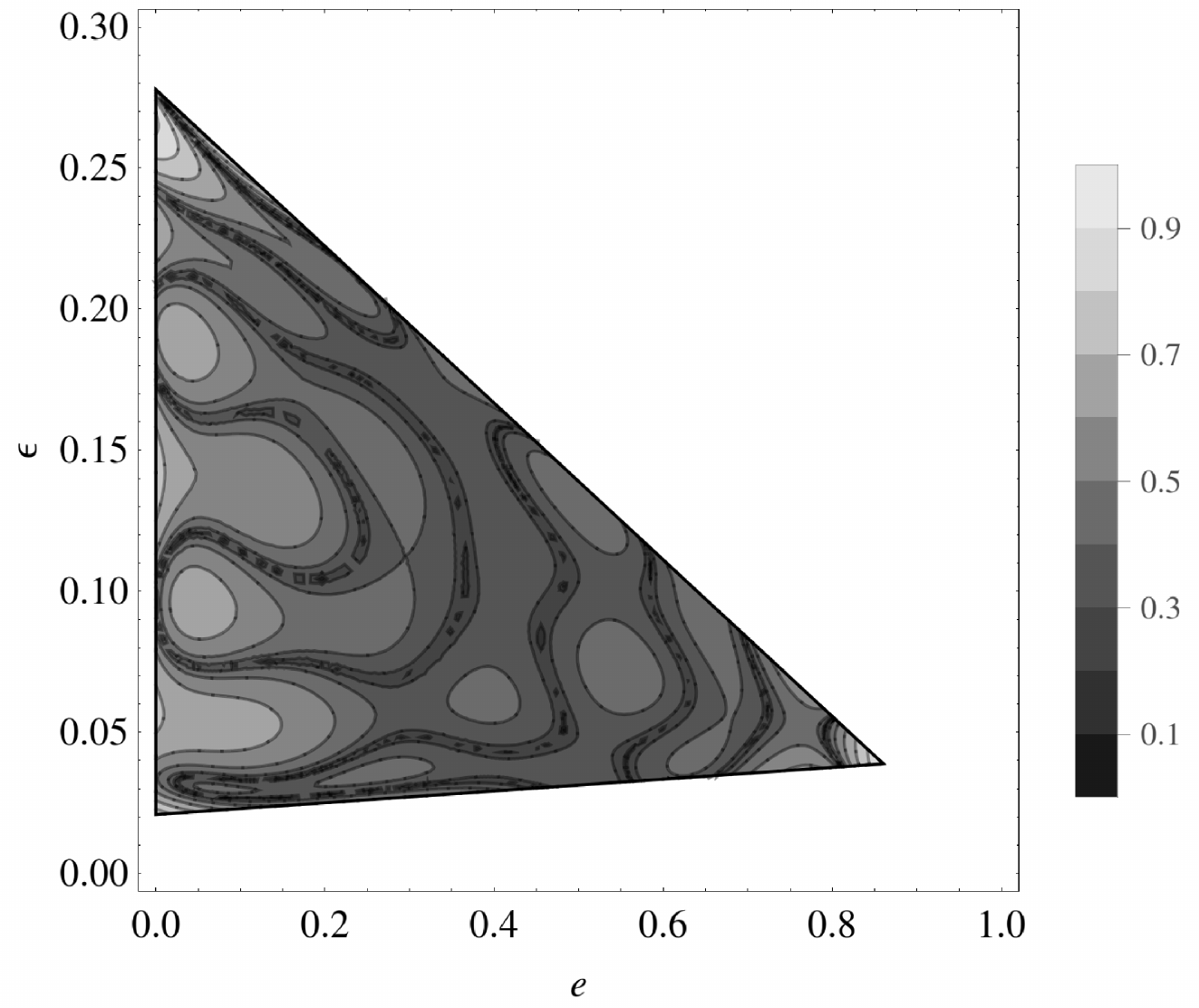}
\mycaption{\label{fig:funkcja_f}
A distribution function $f(e,\epsilon)$ 
consistent with the \rvd profile shown in \figref{fig:RVDprofiles}. The function was obtained with the help of the Keplerian ensemble method, assuming $\mref\!=\!1.8\!\times\!10^{11}\msun$, $Ru_a\!=\!18\,\mathrm{kpc}$ and $Ru_b\!=\!240\,\mathrm{kpc}$. 
The contour plot shows $(f(e,\epsilon)/f_S)^{1/10}$, with $f_S$ being the maximum value of $f(e,\epsilon)$ on the triangular domain.
}
\end{figure}

We model the Galactic potential as consisting of: a disk-like component (accounting for the Galactic rotation curve) and a hot gaseous halo, $\Psi\!=\!\Psi_{\mathrm{disk}}+\Psi_{\mathrm{gas}}$, of total mass $\mref=1.8\!\times\!10^{11}\msun$.
As a starting point for further analyses, we construct an initial \pdf by applying the method of \secref{sec:model} to the \rvd profile in \figref{fig:RVDprofiles}, assuming $M=\mref$,  which is close to the lower bound for this \rvd. In \figref{fig:funkcja_f} we show the resulting \pdf  on the $(e,\epsilon)$ plane. 

\subsection{\label{subsec:simul}Setting the initial conditions}

The first stage toward determining the initial conditions corresponding to the \pdf $f(e,\epsilon)$ shown in \figref{fig:funkcja_f} involves generating a random set of 
initial radii $\mathcal{I}_0=\{u_i\}_{i=1}^N$ in the range $u_a\!<\!u_i\!<\!u_b$ ($N$ is the number of all test bodies),  and with the number density $\nu_u[f]$ from \eqref{funkcja_nu}. 
This task 
of generating random numbers non-uniformly distributed in a range
can be solved as follows. 
Once we find an approximate function 
interpolating an array of $P$ pairs  $\{(u_p,x'_p\!\!\equiv\!\!\nu_{u_p}[f])\}_{p=1}^P$   ordered w.r.t the first argument, we can compute, by numerical integration, an array consisting of triples\smallskip\\ {\centerline{ $\textstyle{\{(x_p\!\!=\!\!\int_{u_a}^{u_p}\nu_{\tilde{u}}[f]\ud{\tilde{u}},\,u_p,\,u'_p\!\!=\!\!(x'_p)^{-1}\!\!=\!\!(\nu_{u_p}[f])^{-1})\}_{p=1}^P}$,}}\smallskip\\ and finally, by a suitable procedure interpolating between the knot points $u_p$, we obtain a smooth function  $u(x)$. 
With a version of a quartic spline interpolation, we can obtain a smooth $u(x)$  with continuous derivatives up to fourth order (in addition to known tabulated $u(x)$ we know also its derivative $u'(x)\!\equiv\! 1/x'(u)$).
Provided that $\nu_{u_p}[f]$ has been normalized to unity,
$\int_{u_a}^{u_b}\nu_{\tilde{u}}[f]\ud{\tilde{u}}\!=\!1$, the smooth $u(x)$  maps  random numbers $x$ uniform in the range (0,1) to random numbers $u$ with the required nonuniform distribution $\nu(u)$ in the range $(u_a,u_b)$.
It may be
instructive to note that $u(x)$ is the inverse of the cumulative probability
function $\chi(u)\!\equiv\!\int_{u_a}^u \nu(\tilde{u})\ud{\tilde{u}}$ of the probability density $\nu(u)$: $x\!=\!\chi(u)$.\footnote{
\label{ft:rnd}With any function $\nu(u)$ positive and integrable to unity on an interval $(u_a,u_b)$  a coordinate change $u\to x=\chi(u)$ can be associated, where $\chi(u)=\int_{u_a}^{u}\nu(\tilde{u})\ud{\tilde{u}}$ is a growing function of $u$, $\chi(u_o)=0$ and $\chi(u_b)=1$.
Let $\chi^{*}$ denote the inverse function of $\chi$: $\chi^*(x)=u$.
From the identity $x\equiv\int_0^x1\ud{\tilde{x}}\equiv\int_{u_a}^{u}\chi'(\tilde{u})\ud{\tilde{u}}\equiv\chi(u)$ it simply follows that if $x$ is uniformly distributed on the unit interval $\br{0,1}$ then $u=\chi^{\star}(x)$ is distributed with the probability density $\nu(u)$ on the interval $(u_a,u_b)$. } 

In the next stage, we need to ensure spherical symmetry  of the initial state. We assign to $\mathcal{I}_0$ a set $\{(\theta_i,\phi_i)\}_{i=1}^N$ of spherical coordinates  of directions uniformly distributed on the unit sphere. From the general result in footnote \ref{ft:rnd} it follows that if $\theta=\arccos\br{1-2x}$ and $\phi=2\pi y$, then $(\theta,\phi)$ will be uniformly distributed on the unit sphere once both $x$ and $y$ are uniformly distributed on independent unit intervals. 
This gives us the initial positions $\mathcal{I}_{1}=\{(u_i,\theta_i,\phi_i)\}_{i=1}^{N}$. 

Next, to obtain initial velocities, we choose random parameters $(e,\epsilon,\psi)$ consistently with the initial \pdf, assigning to each $(u_k,\theta_k,\phi_k)\in\mathcal{I}_{1}$
an elliptic orbit a particular test body would follow in the point mass field.
To this end we consider  triples of random numbers $(e,\epsilon,X)$ 
uniformly distributed in their respective range: $e\!\in\!(0,1)$, $\epsilon\!\in\!\br{0,1/(2u_a)}$ and
$X\!\in\!\br{0,f_S}$, with $X$ being 
an auxiliary variable and $f_S\!=\!\mathrm{max}\{f(e,\epsilon)\!\!: (e,\epsilon)\!\in\! S\}$. For each $u_k\in\mathcal{I}_0$ 
 we carry on generating random triples $(e,\epsilon,X)$ 
until we encounter one (labeled with a subscript $k$) for which 
both $X\!<\!f(e_k,\epsilon_k)$ and $(e_k,\epsilon_k)\!\in\! S(u_k)$. 
This procedure yields a set of random pairs $\mathcal{I}_2=\{(e_i,\epsilon_i)\}_{i=1}^{N}$ with a non-uniform number density distribution $f(e,\epsilon)$ and each confined to a
$u_i$-dependent region $S(u_i)$. To each $u_k$ we also assign  
its respective random angle $\psi_k$ uniformly distributed  in the range $(0,2\pi]$ and fixing the plane of the corresponding ellipse.
In effect, we obtain a set $\mathcal{I}_{2}=\{(e_i,\epsilon_i,\psi_i)\}_{i=1}^{N}$ and form the set $\mathcal{I}=\mathcal{I}_1\oplus\mathcal{I}_2=\{(u_i,\theta_i,\phi_i,e_i,\epsilon_i,\psi_i)\}_{i=1}^{N}$.
Finally, by applying the transformation \eqref{transformacja} to each element of 
$\mathcal{I}$ we obtain the required set of initial positions and velocities in
spherical coordinates, leading to an initial randomly generated \rvd overlapping well  with that in \figref{fig:RVDprofiles} in the region of interest.

\subsection{\label{sec:potential}Gravitational potential}

The $\Psi_{\mathrm{disk}}$  part of $\Psi$ is described by the  thin disk model:
\begin{equation}\label{dysk_potential}
{ \Psi_{\mathrm{disk}}(\rho,\zeta)= -4G\,\int_{0}^{\infty}\mathrm{d}\tilde{\rho}\,\frac{\tilde{\rho}\,K(k)\,\sigma(\tilde{\rho})}{\sqrt{(\rho+\tilde{\rho})^2+\zeta^2}}},
\end{equation}
with $\sigma(\rho)$ being the column mass density of a finite-width disk found by recursions from the Galactic rotation curve  in \mycitep{2014A&A...566A..87J}. Here, $k\!=\!\sqrt{\frac{4\,\rho\,\tilde{\rho}}{(\rho+\tilde{\rho})^2+\zeta^2}}$ and $K$ is the elliptic integral of the first kind defined in \mycitep{Ryzhik}. 
Most of the mass is enclosed within the inner disk $\rho\!<\!20\kpc$: $M_{20}\!=\!1.49\!\times\!10^{11}\msun$, while $M_{30}\!=\!1.51\!\times\!10^{11}\msun$. The outer  $\rho\!>\!30\kpc$ disk's contribution to \eqref{dysk_potential} is thus negligible and we can limit the integration to  $\tilde{\rho}\in(0,30)\kpc$.
To reduce the computation time, we tabulated the integral \eqref{dysk_potential} at mesh points $\{\rho_j,\zeta_k\}$, obtaining a smooth $\Psi_{\mathrm{disk}}$ by means of  the interpolating series $\tilde{\Psi}_{\mathrm{disk}}(\rho,\zeta)\!=\!\sum_{p,q,r}\omega_{pqr}\,\rho^p\zeta^q(\rho^2+\zeta^2)^{-r/2}$ with the coefficients $\omega_{pqr}$ found by the least squares method, minimizing the discrepancy between \eqref{dysk_potential} and the series evaluated at the mesh points. 
Within the desired accuracy, we found this approximation procedure to be numerically more efficient  than the usual two-dimensional interpolation. 

Based on the \oka absorption-line strengths in the spectra of galactic nuclei and galactic sources, \mycitet{2012ApJ...756L...8G} found large amounts of baryonic mass in the form of hot gas surrounding the Galaxy. Assuming a homogeneous sphere model, they found the electron density $n_e$ of  $2.0\!\times\! 10^{-4}/\mathrm{cm}^{3}$ and the path length $L$ of $72\,\mathrm{kpc}$. Among other parameters, the total mass of the gas depends on the gas metalicity and the oxygen-to-helium abundance. For a reasonable set of parameters they found the total mass to be $1.2-6.1\!\times\!10^{10}\msun$. We may assume $M_{\mathrm{gas}}=3.0\!\times\!10^{10}\msun$ consistently with these values. More recently, by applying the same observational method, \mycitet{2013ApJ...770..118M} found the mass function $M(r)$ of the circumgalactic hot gas  using a modified density profile\smallskip\\ 
\centerline{$n(r)={n_0}\br{1+(r/r_c)^2}^{-3\lambda/2}$.}\smallskip\\
We use it as the source of the spherical component $\Psi_{\mathrm{gas}}(r)$,
with the parameters $n_0=0.46\,\mathrm{cm}^{-3}$, $r_c=0.35\,\mathrm{kpc}$ and $\lambda=0.58$  allowable by the best fit to the measurements.
Then the integrated mass  is $M_{\mathrm{gas}}=3\!\times\!10^{10}\msun$ at $r=100\,\mathrm{kpc}$.
For $\lambda\leq1$, the mass function is divergent and the integration must be cutoff at some radius, which is to some extent arbitrary. The cutoff at $100\kpc$  falls within the limits $18\kpc$ and $200\kpc$ on the minimum and maximum mass of the halo   considered in \mycitep{2013ApJ...770..118M}.

\subsection{\label{sec:eqofmot}Numerical solution of the equations of motion}
We consider a test particle  of mass $m$ 
in cylindrical coordinates $(\rho,\varphi,\zeta)$, moving in an axi-symmetric gravitational field described by the potential $\Psi$.  In this symmetry, 
the angular momentum component $J_{\varphi}=m\,\rho^2\d{\varphi}{t}$ is conserved. On account of $\varphi$ being a monotone function of the time $t$
for orbits with $J_{\varphi}\ne0$, we may regard $\varphi$ as the independent parameter. In this parametrization, the Hamilton equations reduce to
\begin{equation}\label{eq.motion}
\begin{aligned}
\d{\rho}{\varphi}&=\frac{\rho ^2}{{\mathcal{J}}} v_\rho, &  \d{v_\rho}{\varphi}&=
\frac{{\mathcal{J}}}{\rho }-
\frac{\rho ^2}{{\mathcal{J}}}\,\pd{{\Psi}(\rho ,\zeta )}{\rho },\\
\d{\zeta}{\varphi}&=\frac{\rho ^2}{{\mathcal{J}}} v_\zeta, & \d{v_\zeta}{\varphi}&=-
\frac{\rho ^2}{{\mathcal{J}}}\,\pd{{\Psi}(\rho ,\zeta )}{\zeta}, 
\end{aligned}
\end{equation}
with $\mathcal{J}\!=\!J_{\varphi}/m$ being the angular momentum per unit mass and $v_{\rho}$ and $v_{\zeta}$ the velocity variables in the orthonormal basis of the coordinate lines $\rho,\zeta$ (see  Appendix \ref{app:eqsderiv} for a
derivation of this from a nonstandard Hamiltonian).
We solve \eqref{eq.motion} numerically by using a $4$-order Runge-Kutta method with adaptive step size. 
controlled so as to keep below some 
small threshold value
the relative change  $\abs{\Delta\mathcal{E}/\mathcal{E}}$  in the energy  per unit mass 
$\mathcal{E}=\frac{1}{2}\br{v_\rho ^2+v_\zeta ^2+\mathcal{J}^2/\rho ^2}+{\Psi}(\rho ,\zeta )$. 
The relative change of energy along each trajectory during our simulation is then always smaller than $10^{-6}$:  $\abs{\mathcal{E}(t)/\mathcal{E}(0)-1}\!<\!10^{-6}$ for all $t$, and this precision suffices for the purposes of this work.

\subsection{\label{sec:results}Results and discussion}

\begin{figure*}
\begin{tabular}{|\spc l\spc|\spc l\spc|\spc l\spc|\spc l\spc|}
\hline
&&&\\
{\tiny A}&\multicolumn{1}{@{}l}{\tiny J}&{\tiny K}&{\tiny T}\\
\includegraphics[width=0.2375\textwidth]{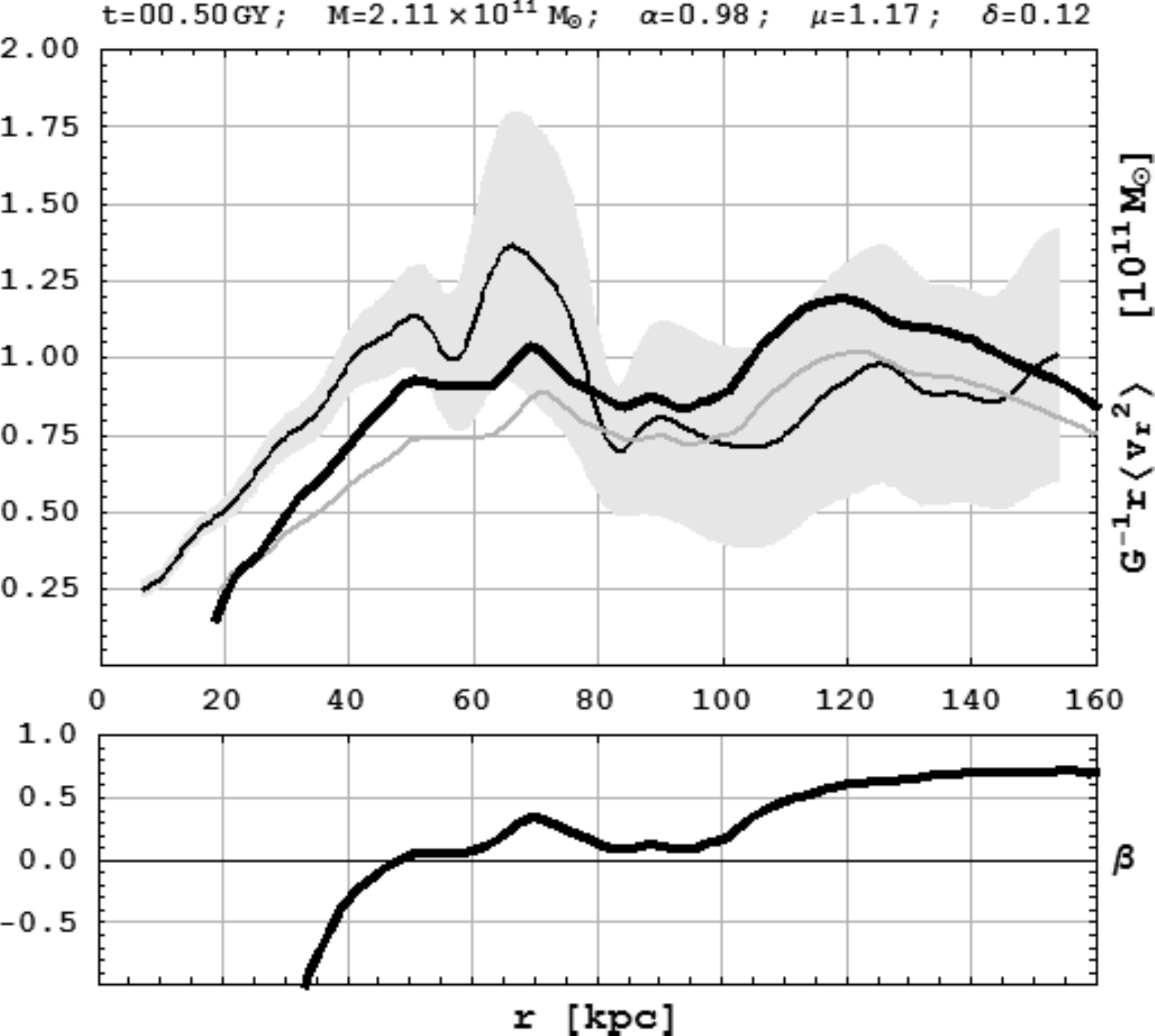}&
\multicolumn{1}{@{}c}{\includegraphics[width=0.2375\textwidth]{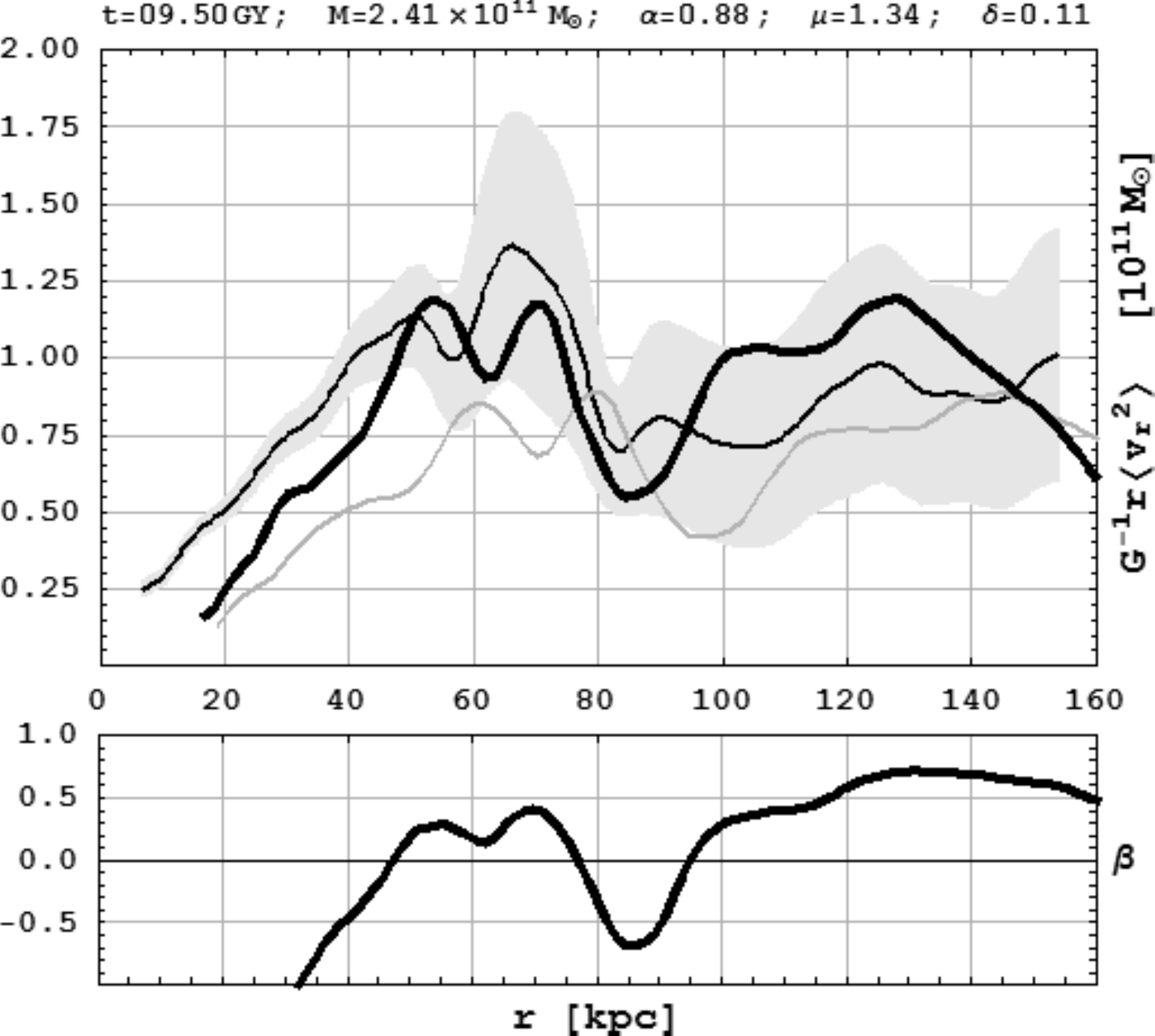}}&
\includegraphics[width=0.2375\textwidth]{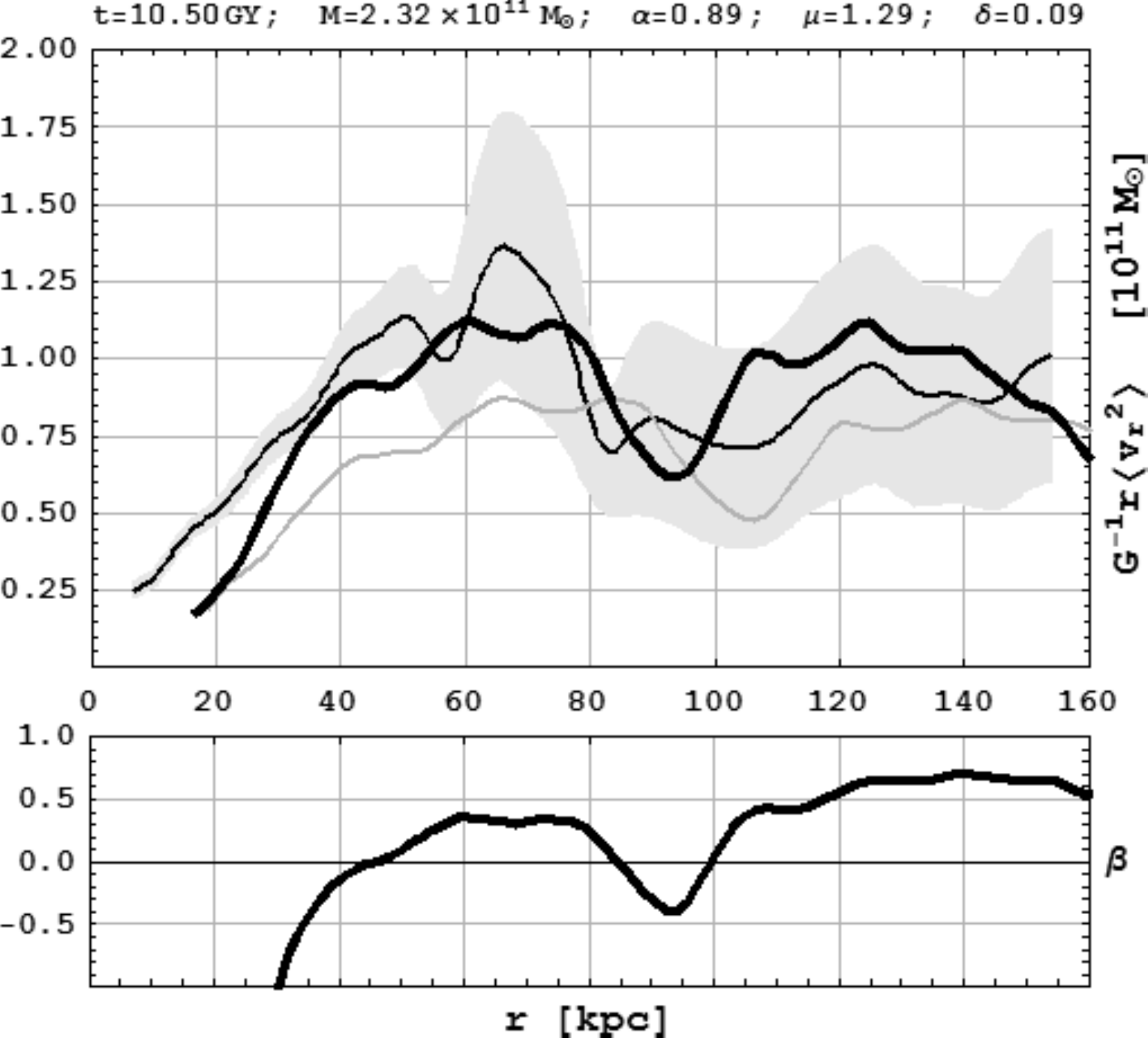}&
\includegraphics[width=0.2375\textwidth]{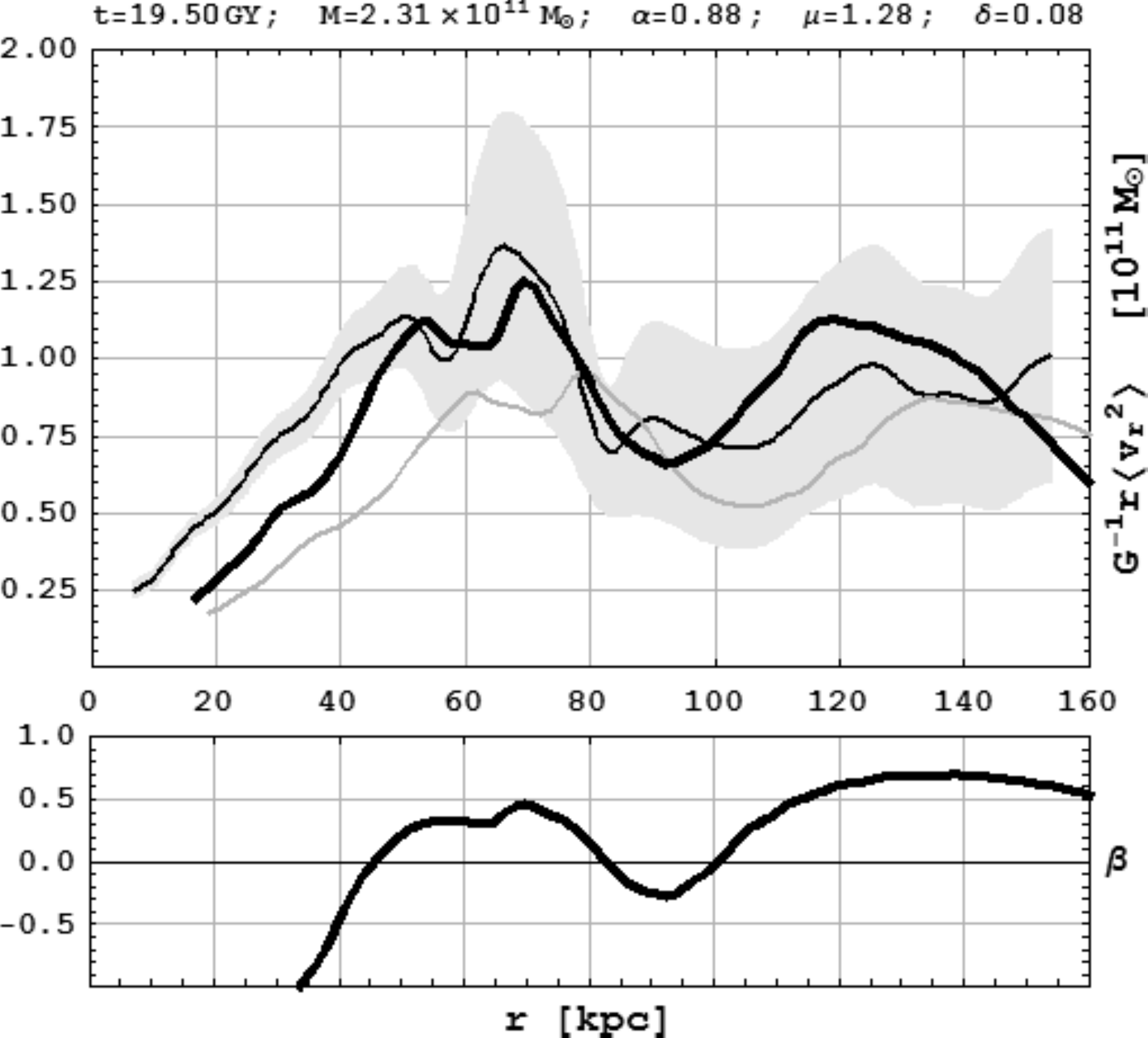}\\
{\tiny B}&{\tiny I}&{\tiny L}&{\tiny S}\\
\includegraphics[width=0.2375\textwidth]{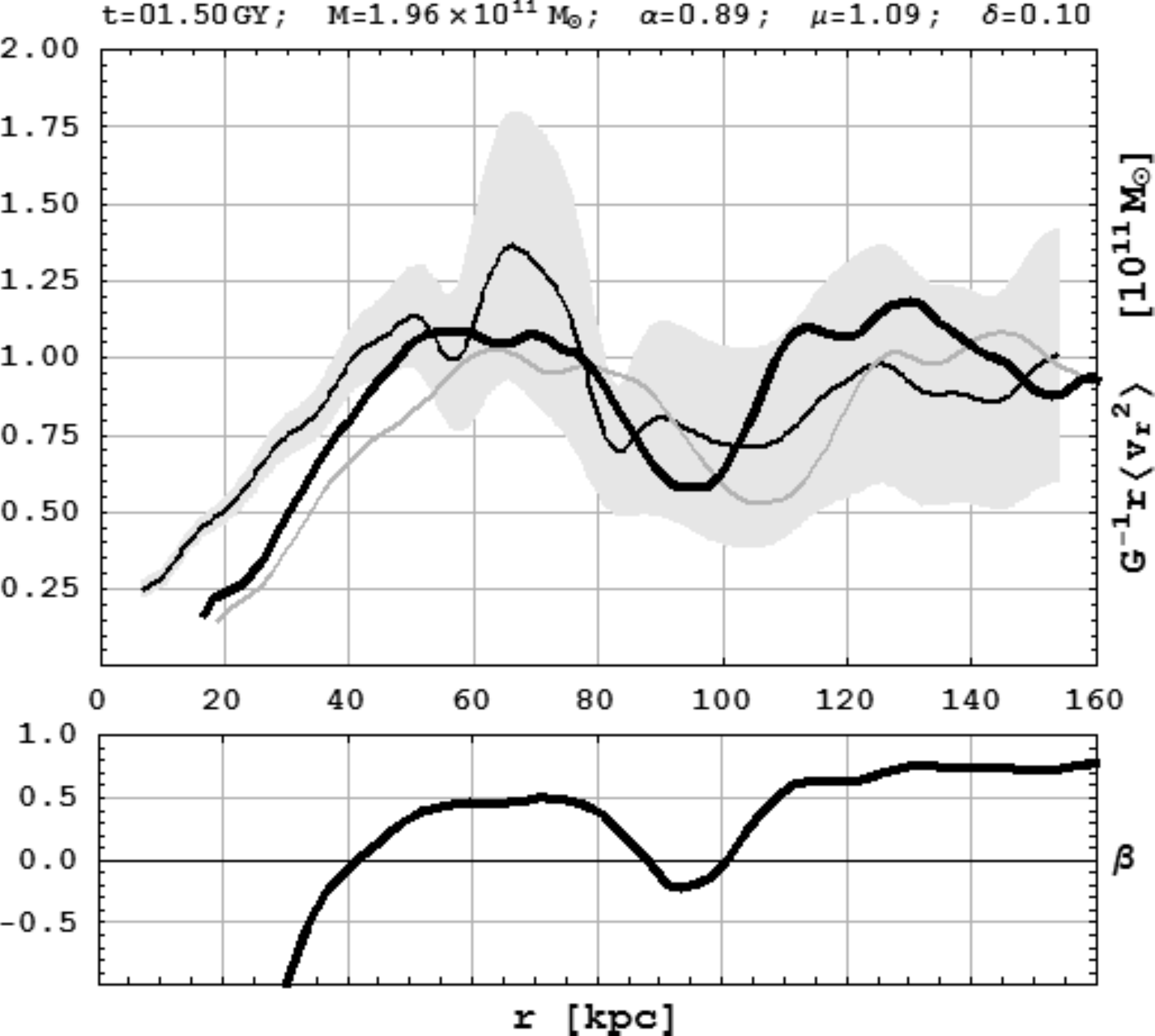}&
\includegraphics[width=0.2375\textwidth]{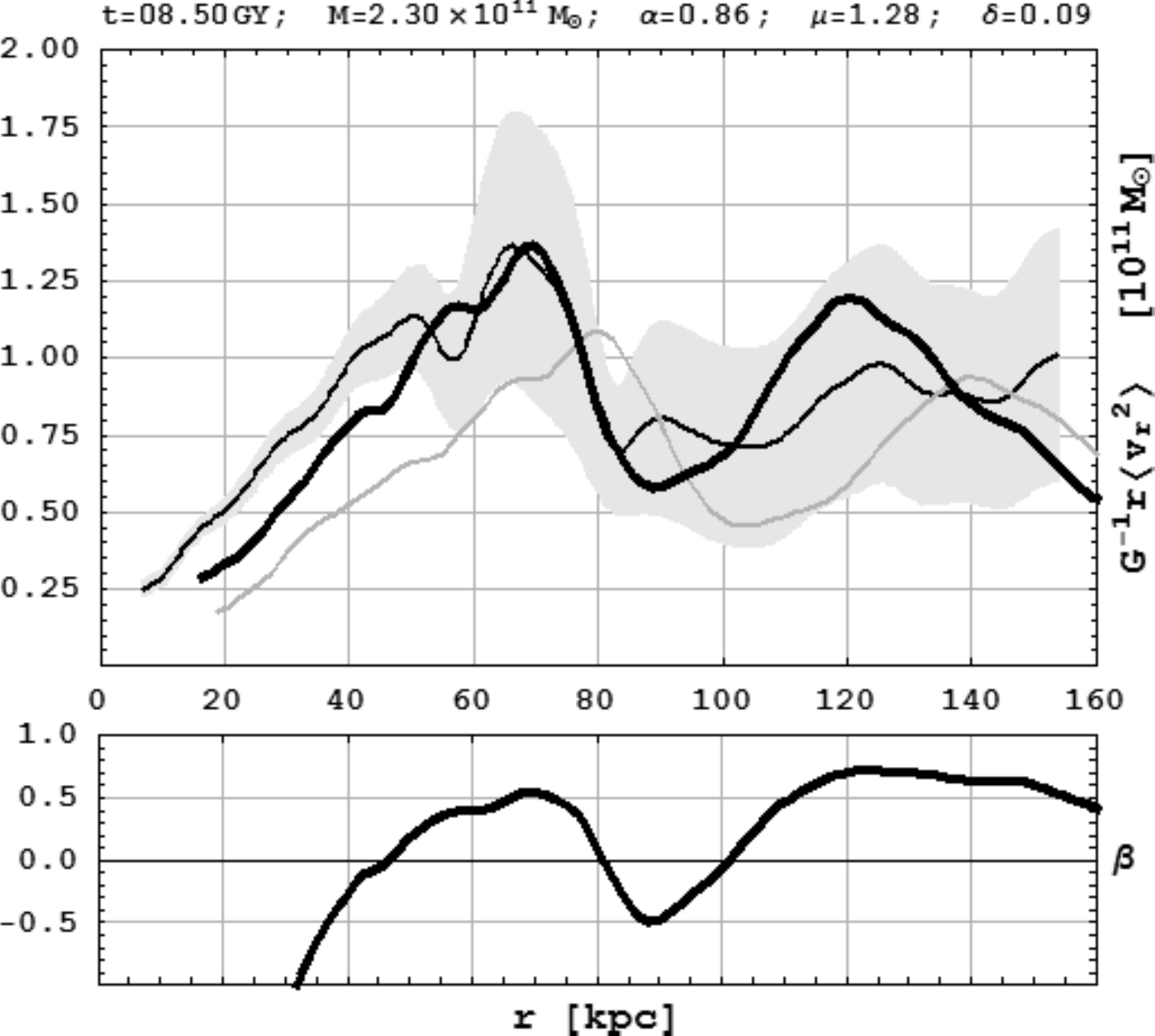}&
\includegraphics[width=0.2375\textwidth]{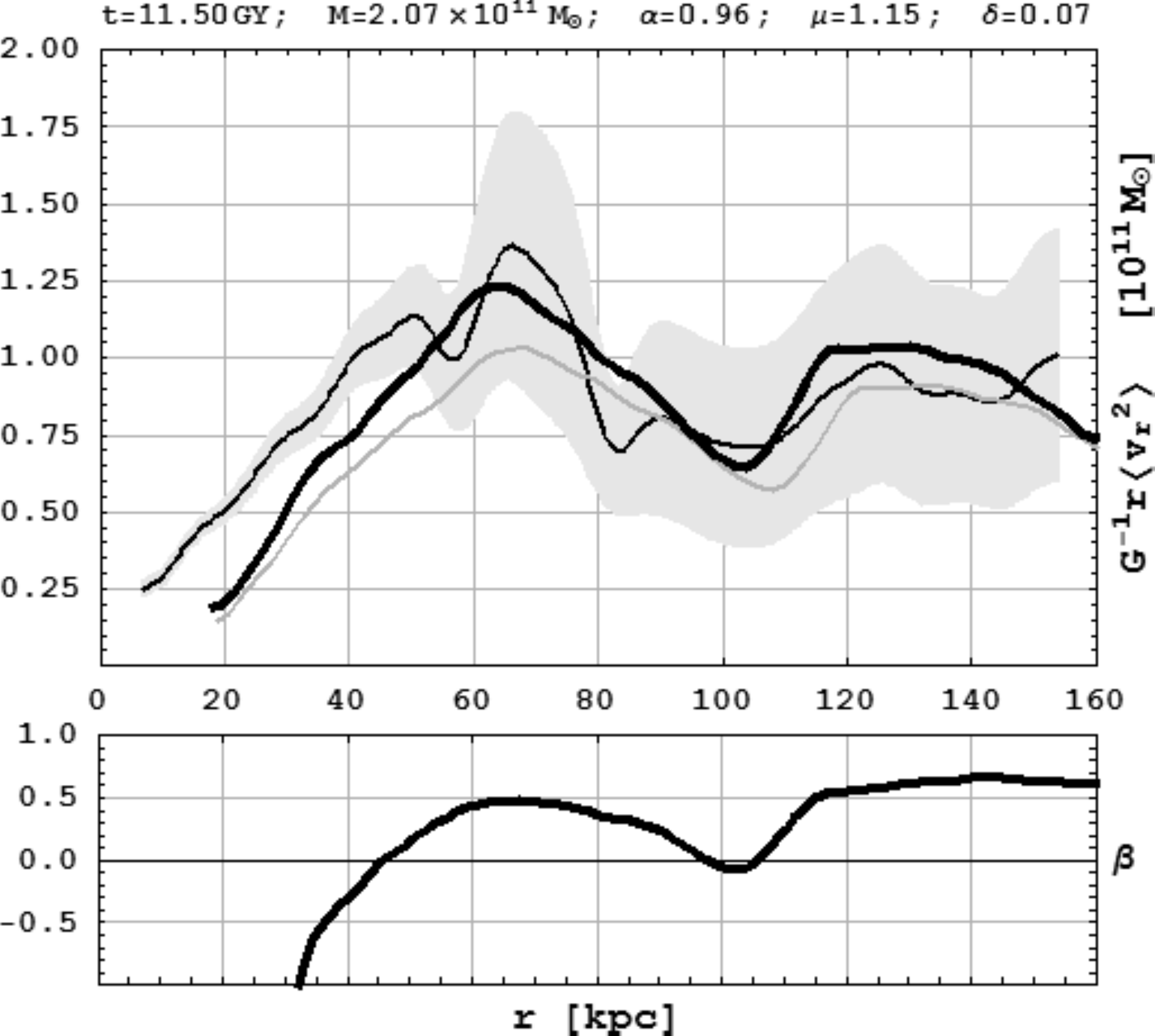}&
\includegraphics[width=0.2375\textwidth]{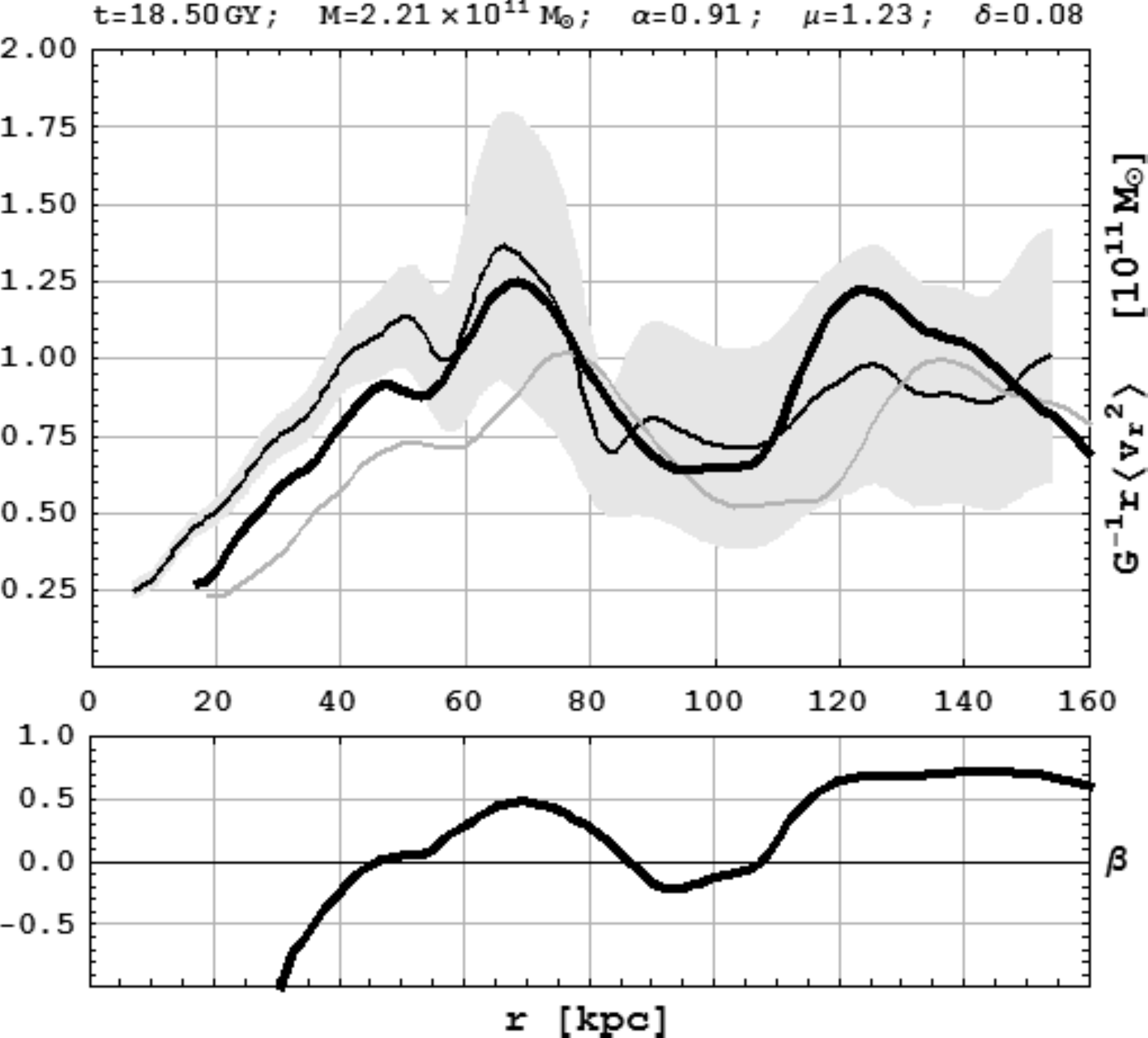}\\
{\tiny C}&{\tiny H}&{\tiny M}&{\tiny R}\\
\includegraphics[width=0.2375\textwidth]{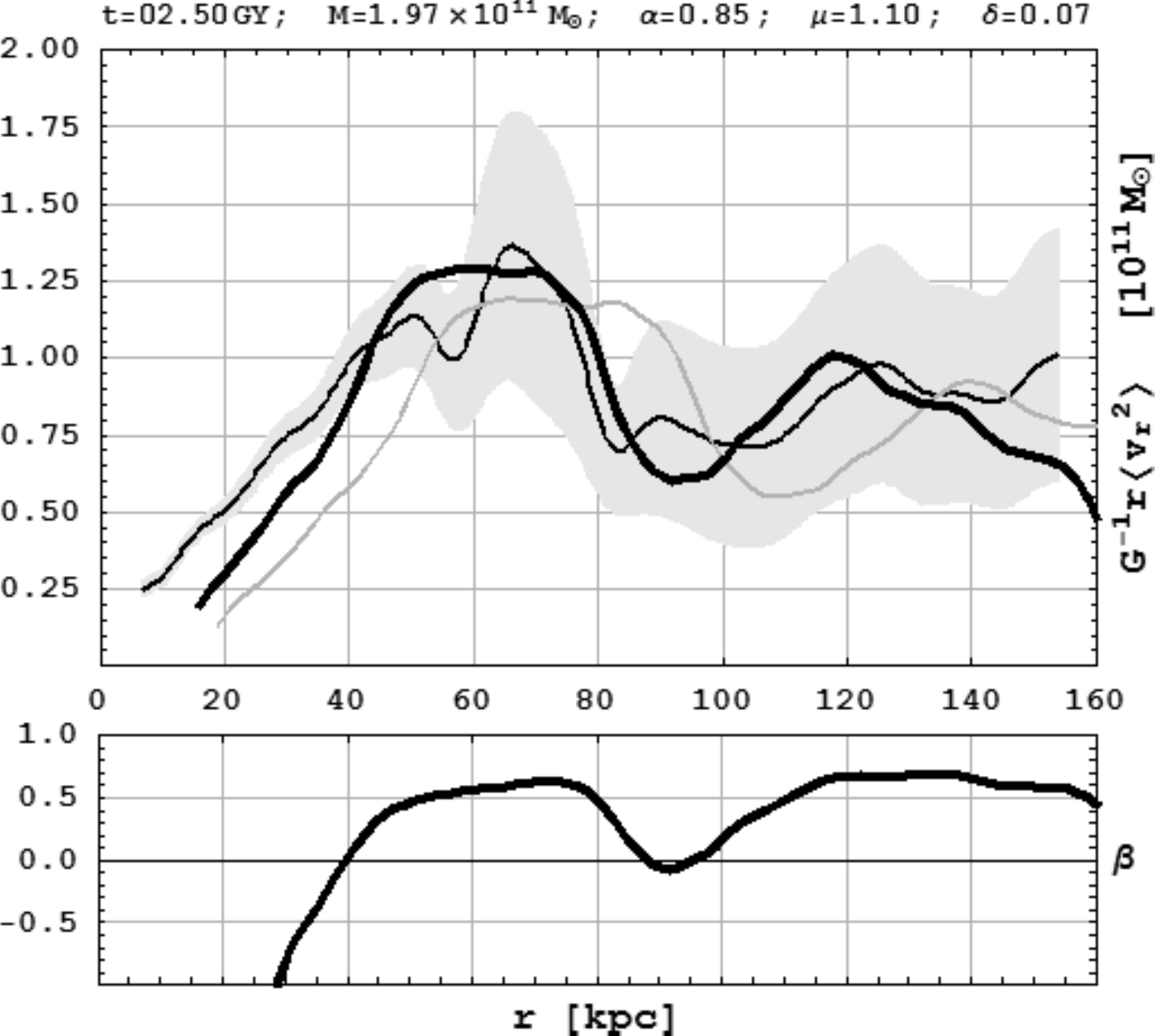}&
\includegraphics[width=0.2375\textwidth]{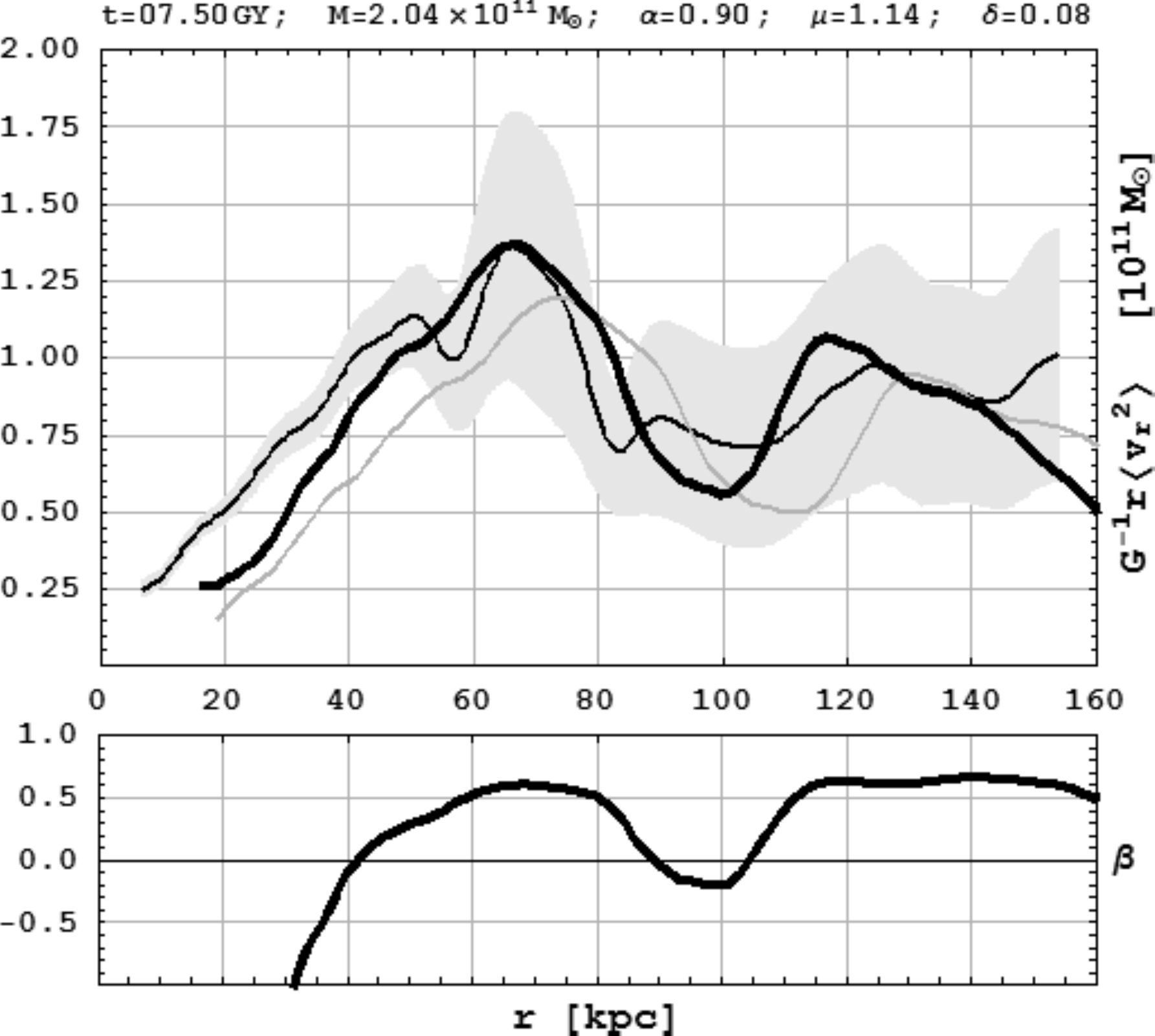}&
\includegraphics[width=0.2375\textwidth]{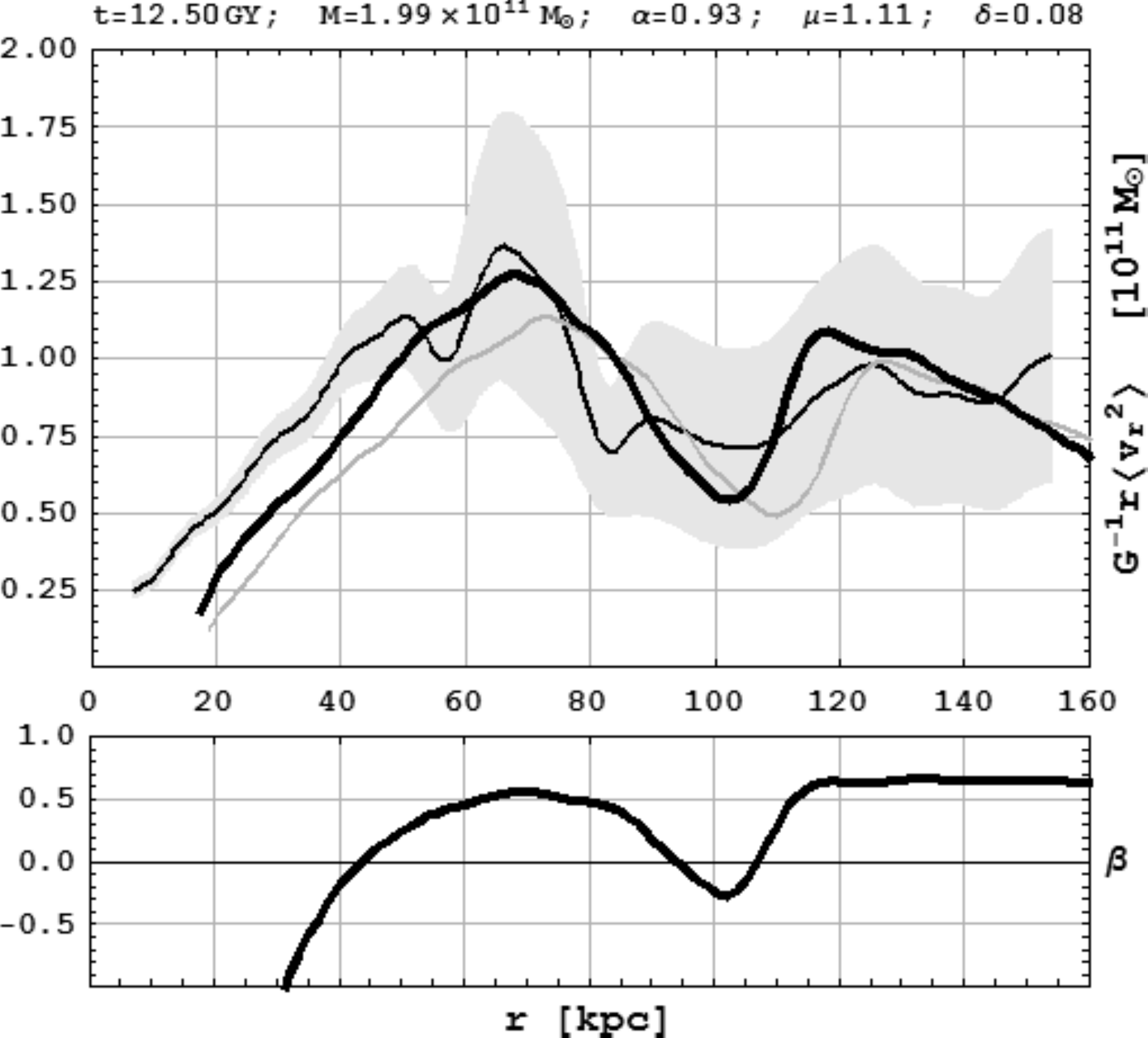}&
\includegraphics[width=0.2375\textwidth]{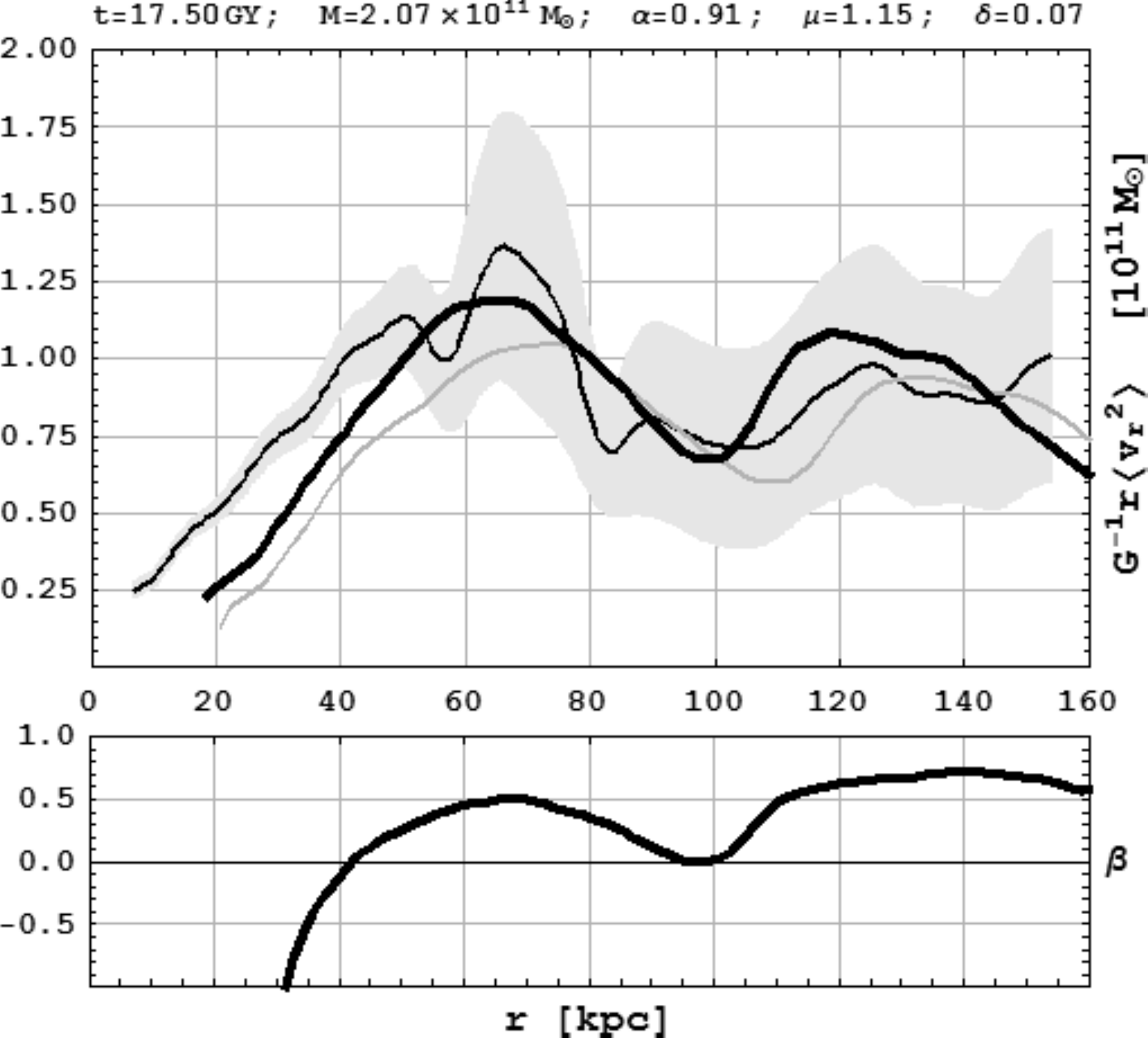}\\
{\tiny D}&{\tiny G}&{\tiny N}&{\tiny Q}\\
\includegraphics[width=0.2375\textwidth]{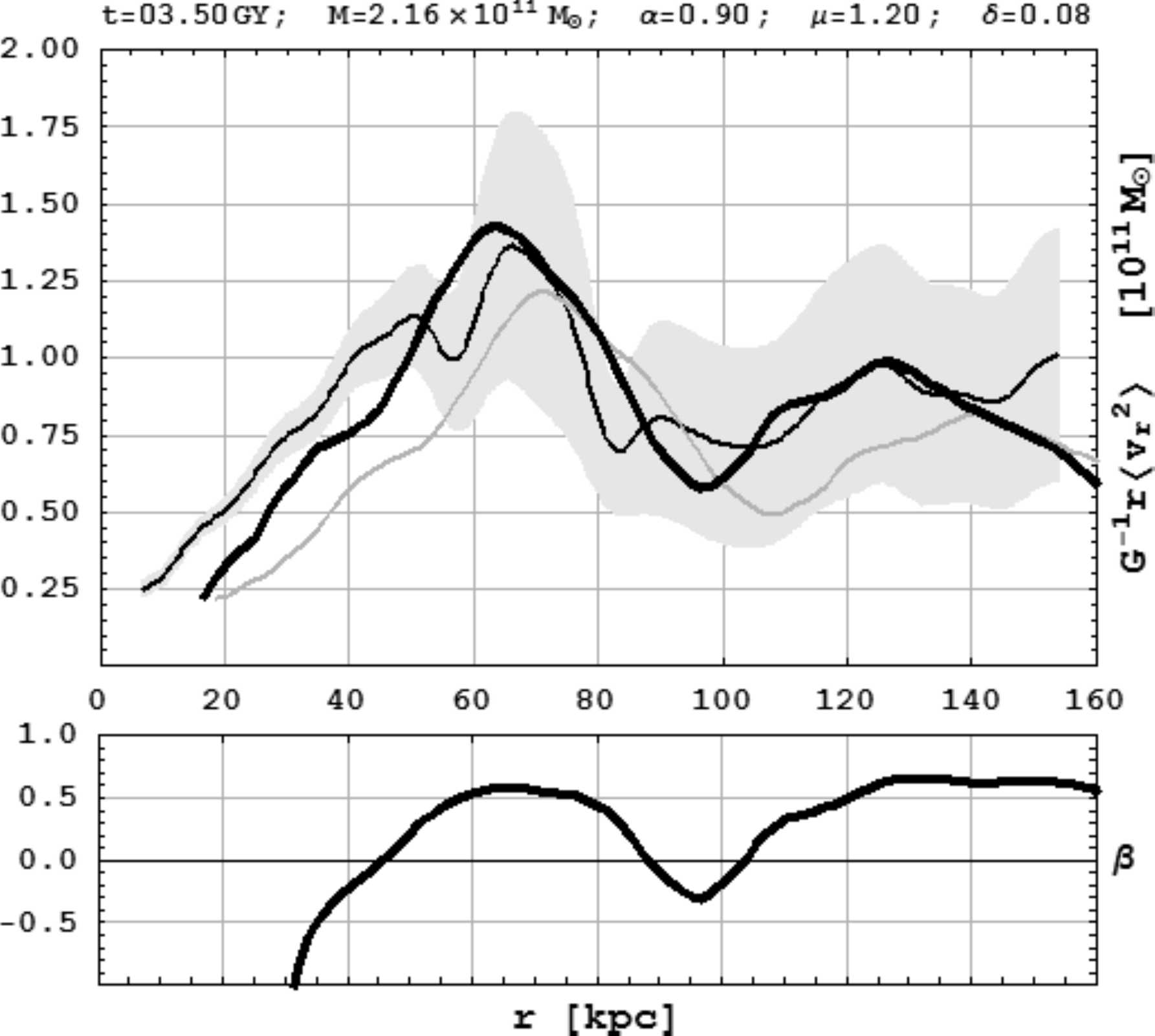}&
\includegraphics[width=0.2375\textwidth]{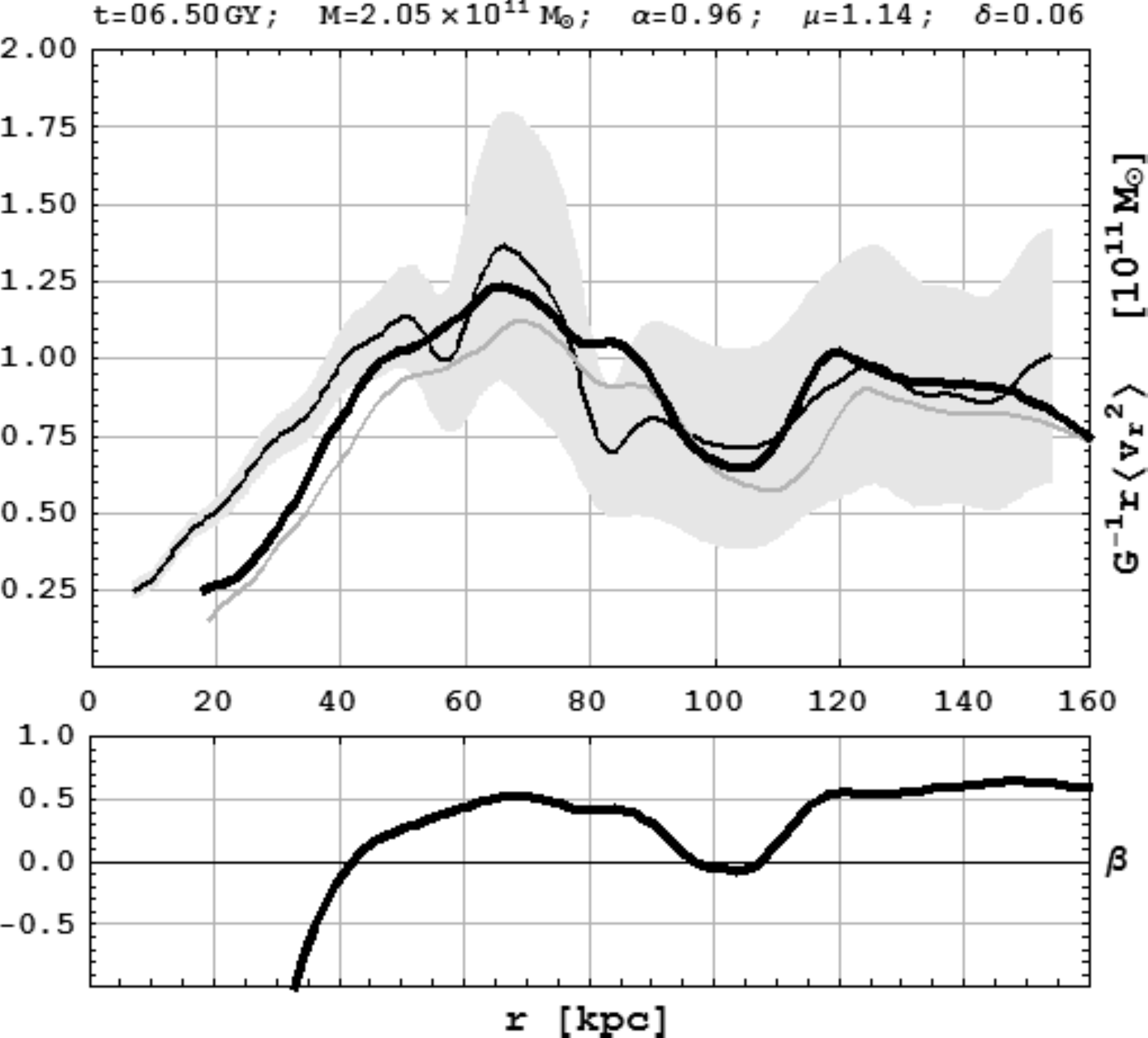}&
\includegraphics[width=0.2375\textwidth]{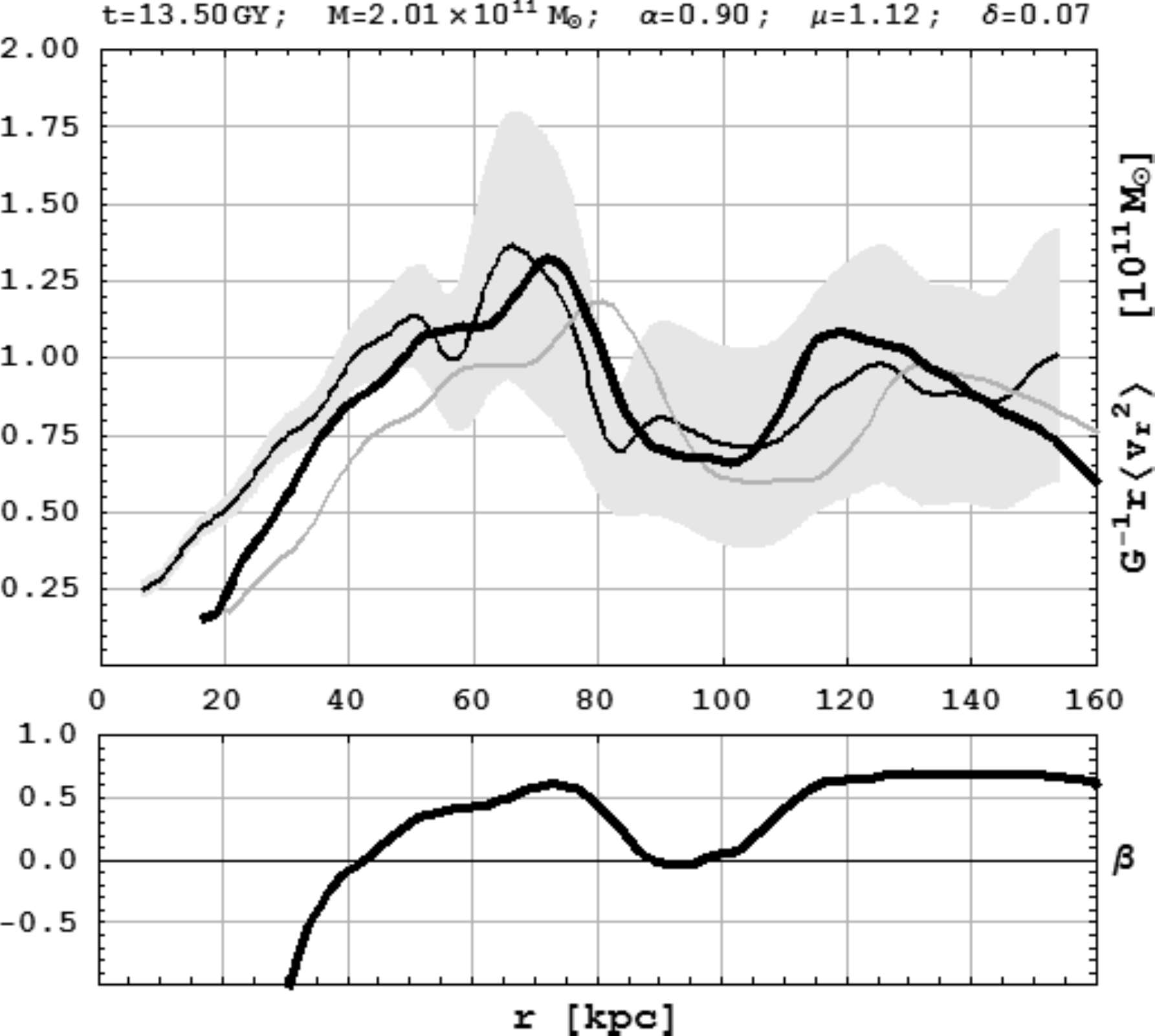}&
\includegraphics[width=0.2375\textwidth]{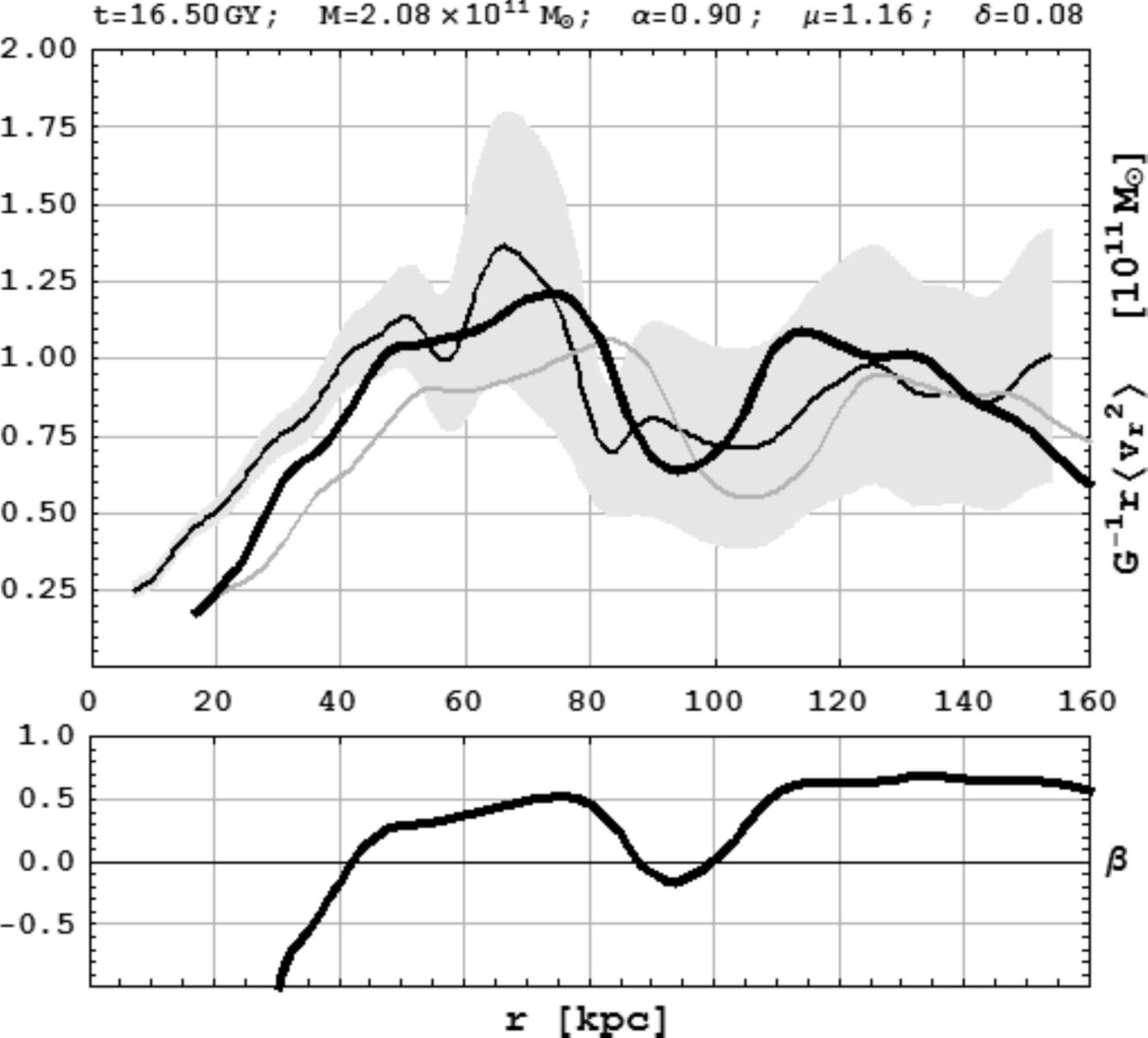}\\ 
\multicolumn{1}{@{}|@{}l}{\,\,\tiny E}&{\tiny F}&\multicolumn{1}{@{}l@{}}{\tiny O}&{\tiny P}\\
\multicolumn{1}{@{}|l@{}}{\!\!\!\;\includegraphics[width=0.2375\textwidth]{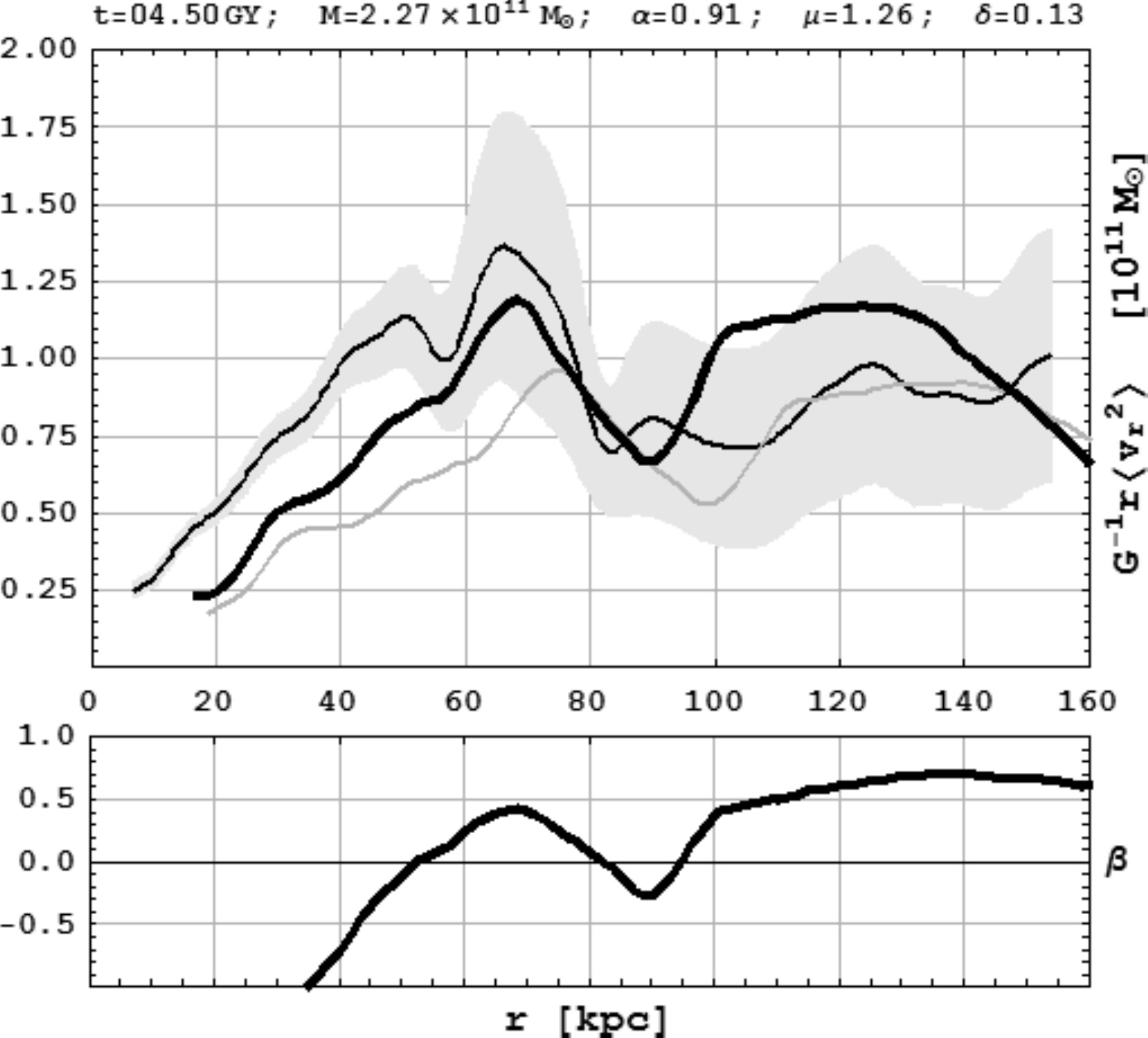}}&
\includegraphics[width=0.2375\textwidth]{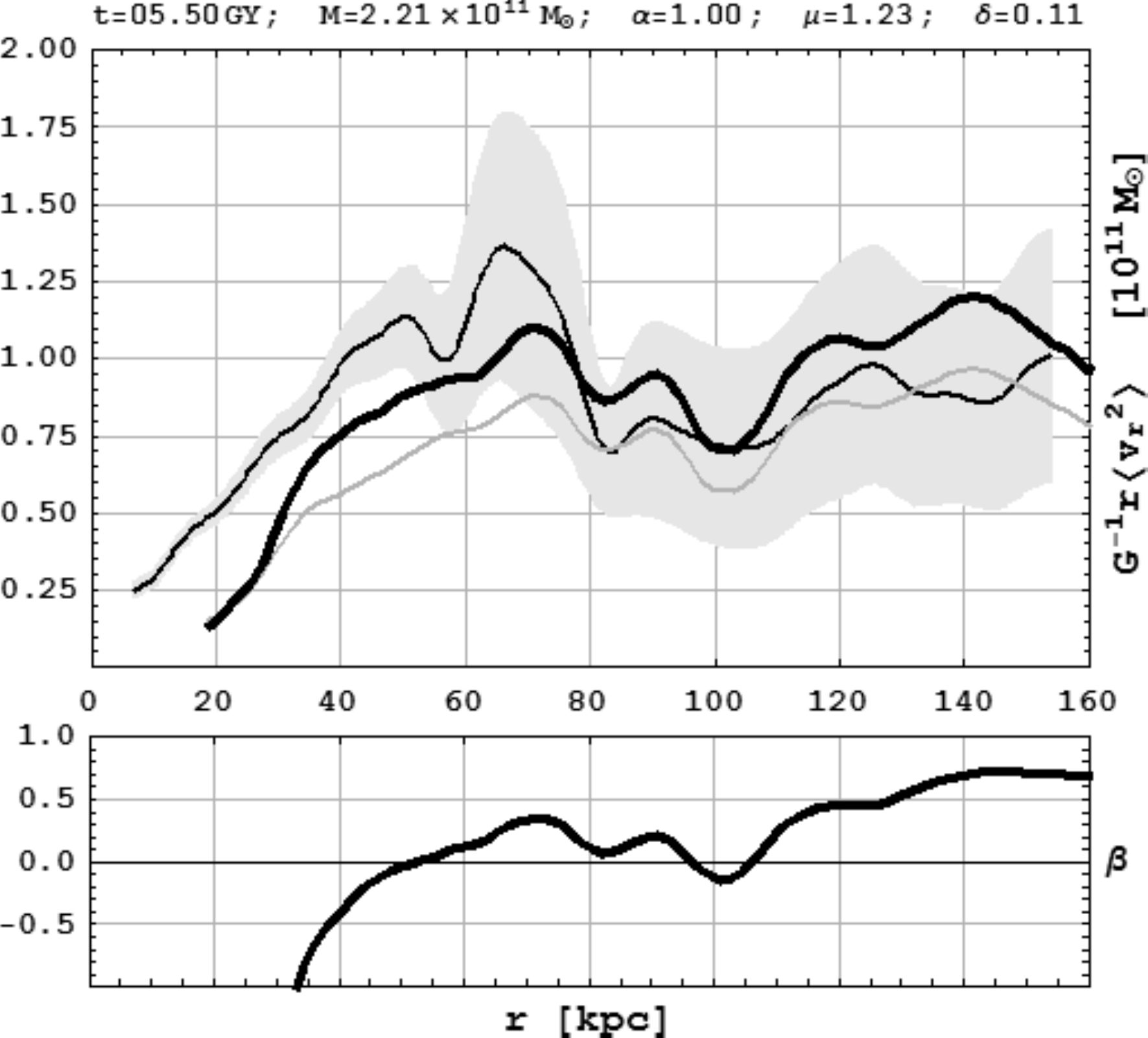}&
\multicolumn{1}{@{}l@{}}{\includegraphics[width=0.2375\textwidth]{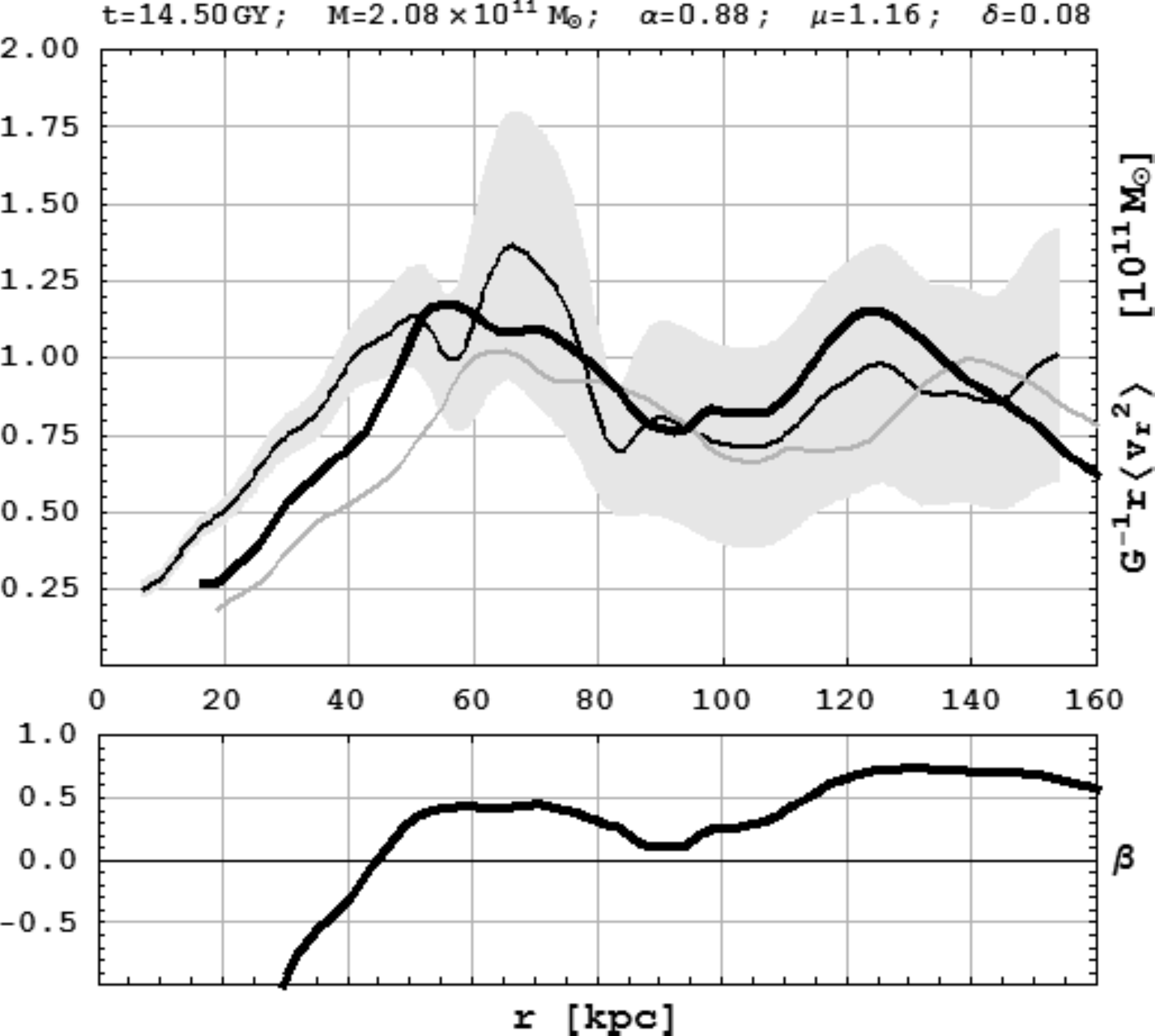}}&
\includegraphics[width=0.2375\textwidth]{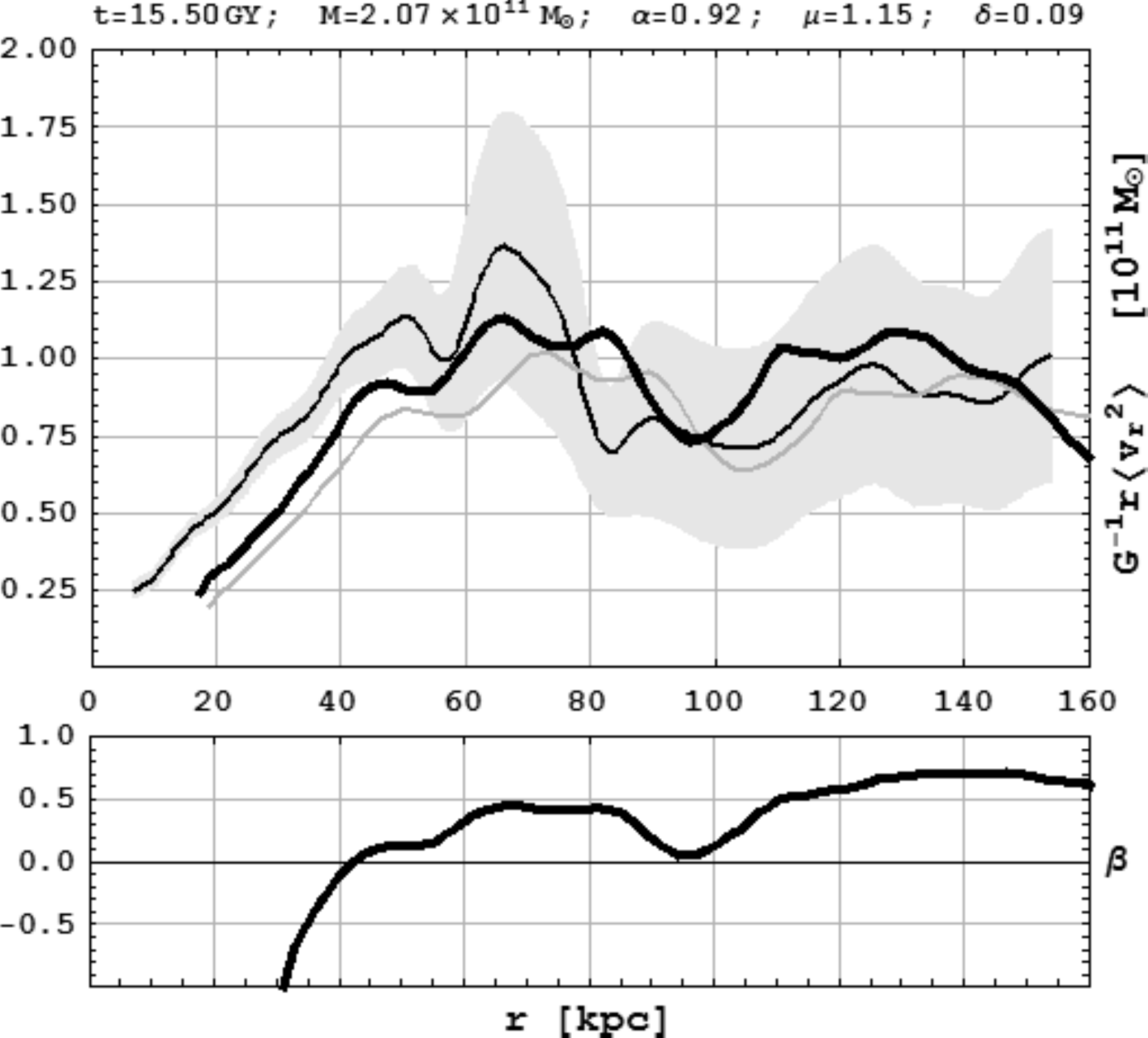}\\
&&&\\
\hline
\end{tabular}
\mycaption{\label{fig:snapshots} A sequence $A,\!B,\dots S,\!T$ of simulated \rvd models \emph{[thick black lines]} shown at distinct simulation instants. The models were obtained from the \emph{[gray line]} \rvd (evolved from the initial \pdf in the $\Psi_{\mathrm{disk}}\!+\!\Psi_{\mathrm{gas}}$ potential of total mass $\mref$) by rescaling 
the horizontal and vertical directions so as to overlap with the background \rvd (from measurements)  \emph{[thin black line]} as good as possible. The \emph{[light gray region]} is the \rvd uncertainty defined by the vertical bars in \figref{fig:RVDprofiles}. The bottom figure in each panel shows a $\beta(r)$ profile corresponding to its respective \rvd model. The decrease in $\beta(r)$ towards negative values in the lower radii region $r<40\kpc$ is a model effect discussed in the text.}
\end{figure*}

Using the numerical procedure of \secref{sec:eqofmot}, we obtained $3665$  trajectories of test bodies starting from the initial conditions of \secref{subsec:simul} and
bound in the potential $\Psi_{\mathrm{disk}}+\Psi_{\mathrm{gas}}$ defined in \secref{sec:potential}. The initial state agrees with the initial \pdf (\figref{fig:funkcja_f}) of a stationary solution of Jeans's problem in the point mass potential and is consistent with the background Galactic \rvd (\figref{fig:RVDprofiles}).  Using these trajectories, we 
determined the \rvd evolution from the initial one and track it through a sequence of snapshots taken at various instants, 
as shown in \figref{fig:snapshots} (with a step size of $\approx1\mathrm{GY}$). 
Each snapshot can be regarded as an independent \rvd model used to estimate the Galaxy mass by comparing the evolved \rvd with the background \rvd. In this approach, the mass we assign to $\Psi_{\mathrm{disk}}\!+\!\Psi_{\mathrm{gas}}$ becomes a function of the simulation time, while the extent of that time has no physical meaning.
 
\indent At this point, it is appropriate to bring  to attention some features of the 
initial \pdf  persisting during the simulation as model effects.
Namely, in the lower radii region, the evolved \rvd values are reduced relative to the background \rvd. The first reason is that
for the initial \pdf from the point mass approximation, a fraction of objects in the modified potential have too high velocities owing to a more extended mass distribution,  and either quickly populate more distant regions or are not bound (in preparing the evolved \rvd profiles we considered only bound trajectories). Consequently, the higher velocity values do not contribute in this region and the \rvd values get reduced. The other model effect is due to a cutoff in the \pdf domain introduced in the Keplerian ensemble method to automatically prevent test bodies from penetrating the interior of a central spherical region where the point mass approximation is violated. 
As so, there is no limit on the number of  almost nearly circular orbits 
the external neighborhood of this region can accommodate -- too much elongated orbits cannot occur there for geometrical reasons,  while the admissible elongated orbits enter this region with their pericentric sides only (where radial motions are almost vanishing).  In consequence of this, 
the overall mean \rvd in this neighborhood gets reduced, below the observed values. Because the initial \pdf has been identified with the \pdf of the Keplerian ensemble, a qualitatively similar reduction mechanism in the evolved \rvd comes about in the modified potential, reflecting in the $\beta(r)$ reduced toward more negative values  in the lower radii region (however, more circular motions in this region could be interpreted as consistent with a contribution from a cold disk). With a better initial \pdf this model effect could be eliminated but it seems of no importance for the accuracy of the total mass determination for which the region of greater radii is more important. A similar cutoff mechanism may increase the number density in the neighborhood of the upper boundary (which we assumed to be of $240\kpc$). 
Namely,  the high \rvd values observed for moderate radii and modeled in the point mass approximation by more elongated elliptic orbits, are reduced again to zero close to the upper boundary. This reduction in the \rvd appears naturally -- the elongated orbits enter this region with their apocentric sides where radial motions are almost vanishing and where test bodies spent a relatively longer time, and this effect can be amplified by increasing the number density of bodies on more circular orbits. Because the model \rvd is compared with the measurements at moderate radii, this effect can be again neglected.

Now, let us come back to the main issue. As mentioned earlier, we want to verify the expectation that the evolved \pdf should be in a sense close to the initial \pdf,  independently of the simulation instant if the point mass approximation well describes the real situation at higher distances.
If the \rvd evolved in the modified potential  turned out to be collapsing to much smaller values or change its shape completely, then this would mean that the mass estimate based on the initial \pdf was wrong and inconsistent with the new evolved \pdf. 

As seen in \figref{fig:snapshots}, although the evolved \rvd changes with the simulation time -- it decreases a little 
in some region and later it grows again - it generally remains high in the larger radii region.  Similarly, the characteristic maxima in the initial \rvd are not destroyed but oscillate. 
\begin{figure}[h]
\centering
\includegraphics[trim = 22pt 12pt 22pt 10pt,
clip,
width=0.46\textwidth]{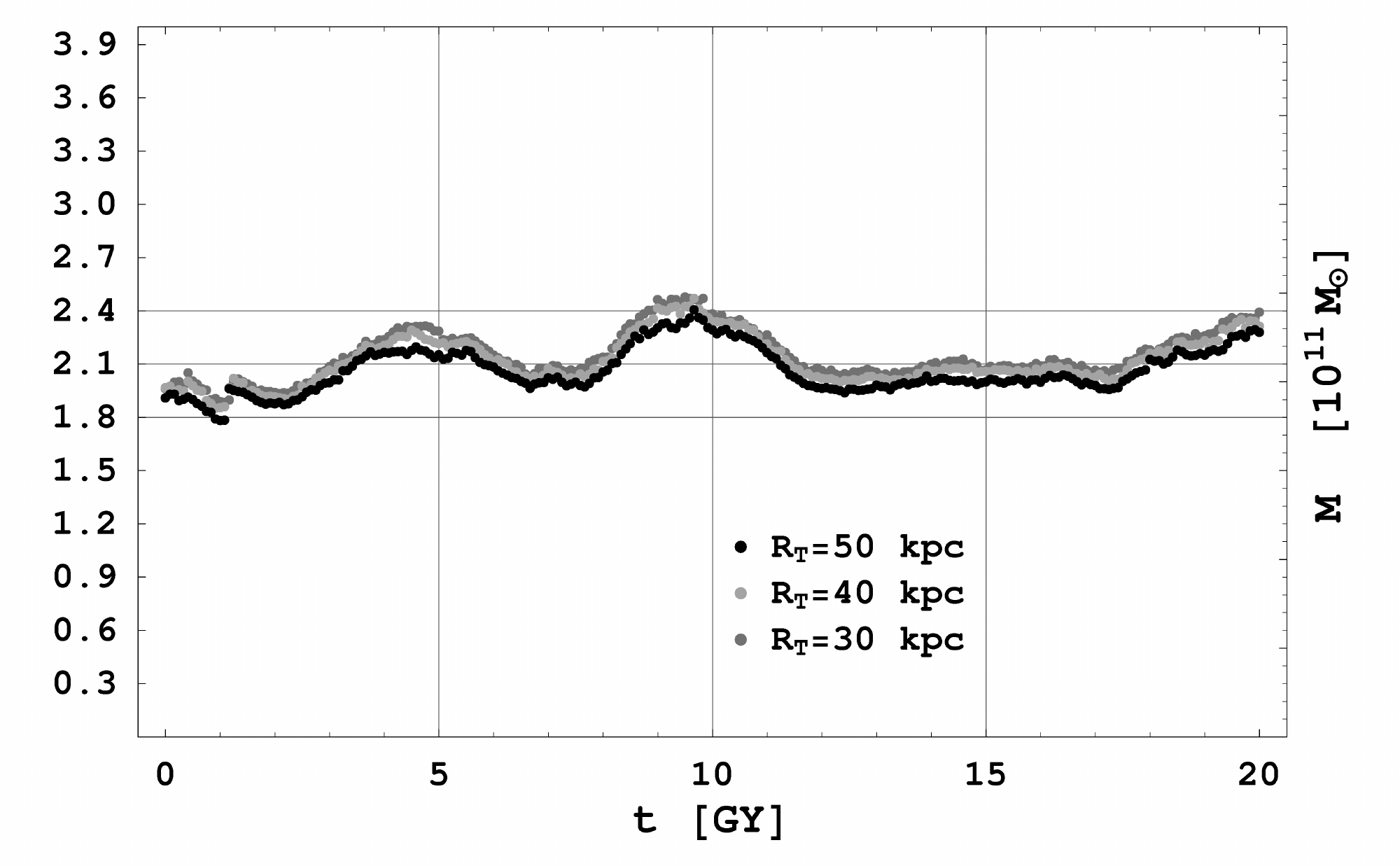}
\mycaption{\label{fig:massf}
Mass estimator $\mu\,\mref$ as a function of the simulation time, with factors $\mu$ obtained from best fit model \rvd at various threshold radii $R_T$. }   
\end{figure}
Besides the evolved \rvd profile (gray line) corresponding to the mass
 $\mref$, in each snapshot there is also shown  a corrected \rvd (thick black-line) which we consider as our model \rvd profile, obtained by multiplying $\mref$ and the radial variable with suitable factors close to unity, respectively $\mu$ and $\alpha$, so as to make the model \rvd coincide with the background one as good as possible in the sense of the least squares.\footnote{{that is, by minimizing  $\delta\!\equiv\!\mref^{-1}[{N}^{-1}\!\sum_{i=1}^{N}\!\br{\mu\!\cdot\!\! Y(r_i/\alpha)-Y_{o}(r_i)}^2]^{1/2}$, where $Y(r)$ is the gray line \rvd, $Y_o(r)$ is the background \rvd, and  the summation is taken over $r_i>R_T$, with a threshold radius $R_T= 40\kpc$ delineating
the less important lower radii region.}} During the simulation run, the length factor $\alpha$ varied in the range $(0.85;\,1.02)$ with the mean $0.92\!\pm\!0.03$,  while the mass factor $\mu$ varied in the range $(1.03;\!1.37)$ with the mean $1.18\!\pm\!0.07$ (see \figref{fig:massf}).  
This gives the total mass estimate of $(2.12\!\pm\!0.13)\!\times\!10^{11}\msun$, oscillating in the range $(1.85;\!2.47)\!\times\!10^{11}\msun$. 

\section{Conclusions}

The lower bound for the Galaxy mass of $\approx2.1\!\times\!10^{11}\msun$ obtained within  the Keplerian ensemble framework suffices to retain, during a numerical simulation run in a modified potential, the qualitative features of the evolved \rvd profile and its values, consistently with the \rvd from halo measurements within $150\kpc$. In this sense the evolved \rvd is stable. These results also substantiate structural stability of the point mass approximation, showing that the lower range for Galaxy mass estimates assuming more general unconstrained \pdf's in this approximation is reliable. 
A possible correction factor $1.16$ to account for the $4$ halo objects rejected in \secref{sec:measurements}, would give a value of $(2.5\pm0.2)\times10^{11}$, consistently with earlier estimates of $2.4\!\times\!10^{11}\msun$ \citep{1992AJ....103.1552M,1987ApJ...320..493L} or with the value of $(2.6\!-\!2.7)\!\times\!10^{11}\msun$ recently inferred from the kinematics of the Orphan stream \citep{2010ApJ...711...32N,2013ApJ...776...26S}  within $\sim60\kpc$.

The crucial role in our analysis is played by the general unconstrained phase space. We stress that the phase space model is not less important than the mass model, and focusing more  attention on generic phase spaces might help to reduce the missing mass problem.

We are aware of that the model potential considered here neglects the 
nonbaryonic dark matter (NDM)
halo thought to extend to very large distances. However, this is the most hypothetical and less constrained Galactic component. As we mentioned in \secref{sec:intro}, the Galaxy mass estimates in the literature differ between each other by a factor larger than two. We presented here the extreme example of a model without NDM halo, and have shown the model to be stable in a sense that it accounts for the measured \rvd at each simulation instant. This shows that our model can be thought of as a collision-less system close to a steady-state. 
The possibility of accounting for the \rvd observations without NDM halo  shows that either the halo is not necessary for understanding motions of the kinematical tracers, or that other observational features (e.g. the measurements of the $\beta(r)$ function) are needed in order to define constraints on the phase space which would allow to disambiguate between various halo mass profiles.  

\medskip
\begin{center} {\large\bf APPENDIX} \end{center}
\renewcommand\thesection{\Alph{section}}
\renewcommand\thesubsection{\thesection.\arabic{subsection}}
\setcounter{section}{0} 

\section{Derivation of a relation between the observable $\avg{\tilde{v}_r^2}$ and the radial dispersion $\avg{v_r^2}$}\label{app1}
Let 
$\vec{v}$ be the Galacto-centric velocity vector. Expressed in terms of its~radial $V_r$ and
transversal  components $V_{\theta}, V_\phi$ it reads $\vec{v}\!=\left[
V_r\sin{\theta}\cos{\phi}\!+\! V_\theta\cos{\theta}\cos{\phi}\! -\! V_\phi\sin{\phi},\right.$\\$\left.
\br{V_r\sin{\theta} + V_\theta\cos{\theta}} \sin{\phi} + V_\phi\cos{\phi},
\,V_r\cos{\theta} - V_\theta\sin{\theta}
\right]$. As
$\vec{e}_r\!=\!\frac{\vec{r}}{\abs{\vec{r}}}\!=\!\sq{\sin{\theta}\cos{\phi},\sin{\theta}\sin{\phi},\cos{\theta}}$ and $\vec{r}_\odot=\sq{\rsun,0,0}$ the l.o.s versor is
$\vec{e}_\rho\!\equiv\!\frac{\vec{r}-\vec{r}_\odot}{\abs{\vec{r}-\vec{r}_\odot}}\!=\!{\textstyle \frac{\sq{r\sin{\theta}\cos{\phi}-\rsun,r\sin{\theta}\sin{\phi},r\cos{\theta}}}{\sqrt{
r^2+\rsun^2-2r\rsun\sin{\theta}\cos{\phi}}}}$. 
We take the mean values $\avg{v_r^2}\equiv\avg{\br{\vec{e}_{r}\circ\vec{v}}^2}$ and $\avg{\tilde{v}_r^2}\equiv\avg{\br{\vec{e}_{\rho}\circ\vec{v}}^2}$ over thin  spherical shells and consider them as functions of the Galacto-centric distance $r$.  
For a spherically symmetric system we define $\avg{\tilde{v}_r^2(r)}$ as
$$\avg{\tilde{v}_r^2(r)}=\frac{1}{4\pi\,\nu(r)}\int_{0}^{\pi}\!\!\!\ud{\theta}\sin{\theta}\int_{0}^{2\pi}\!\!\!\!\ud{\phi}\,
\avg{\br{\vec{e}_\rho(r,\theta,\phi)\circ\vec{v}}^2}_{\mathrm{int}}.$$
Here, $\avg{\cdot}_{\mathrm{int}}$ is the averaging over the velocities weighted by a spherically symmetric \pdf  $f(r,\vec{v}(r))$, normalized so as $\nu(r)\avg{(\cdot)}_{int}\equiv\int (\cdot)f(r,\vec{v}(r))\ud^3{}\vec{v}$, with $\nu$ denoting the number density.  
The scalar product squared $\br{\vec{e}_\rho(r,\theta,\phi)\circ\vec{v}}^2$ is a homogenuous form of second degree in the velocities $V_r,V_\theta,V_\phi$ with coefficients being functions of $r,\theta,\phi$. 
By a direct inspection one can notice that the integration over $\theta,\phi$ of the coefficients standing at  $V_r V_\theta$, $V_\theta V_\phi$, $V_\phi V_r$, gives zero (the velocity products are independent of $\theta,\phi$). 
Thus, upon integration over the velocities, we can focus only on the  terms involving dispersions $\avg{V_r^2}(r)$, $\avg{V_\theta^2}(r)$, $\avg{V_\phi^2}(r)$.  Furthermore, it also follows from the spherical symmetry that $\avg{V_\phi^2}(r)=\avg{V_\theta^2}(r)$ and, trivially, that the ratios  $\avg{V_\theta^2}/\avg{V_r^2}$, $\avg{V_\phi^2}/\avg{V_r^2}$  define the same function of $r$. In accordance with the common convention in the theory of spherical Jeans equations, we express this function in terms of the flattening of the dispersion ellipsoid, $\beta(r)$. Then, by making the substitution  $\avg{V_\theta^2}(r)=(1-\beta(r))\avg{V_r^2}(r)$ and $\avg{V_\phi^2}(r)=(1-\beta(r))\avg{V_r^2}(r)$, we obtain that $$\avg{\tilde{v}_r^2}(r)=\avg{{v}_r^2}(r)\br{1-\frac{\beta(r)}{1+r^2/\rsun^2}\cdot \mathcal{I}(\alpha(r))},$$ where  $\alpha(r)\equiv\frac{2r\rsun}{r^2+\rsun^2}<1$ for $r\neq\rsun$ and
$$\mathcal{I}(\alpha)=\frac{1}{4\pi}\int_{0}^{\pi}\!\!\!\ud{\theta}\int_{0}^{2\pi}\!\!\!\ud{\phi}\,\,\frac{\sin{\theta}\br{\cos^2{\theta}\cos^2{\phi}
+\sin^2{\phi}}}{1-\alpha\sin{\theta}\cos{\phi}}.$$ We recall that all integrals that are zero by symmetries have been already omitted in the  expression for $\mathcal{I}(\alpha)$. On account that the requirements for the integration of a functional series term by term and its limit are met for $0\!\leq \alpha\!<1$,
the integral $\mathcal{I}(\alpha)$ can be calculated by a Taylor series expansion in  $\alpha$  (note that owing to the vanishing of the integrals $\int_0^{2\pi}\!\cos^{m}\phi\ud{\phi}$ with odd $m$, only even powers of $\alpha$ are present in the series). On reducing the summands with the help of the Pythagorean trigonometric identity, the remaining nonzero coefficients in the power series in $\alpha$ arrange  to products of elementary definite integrals
$\mathcal{I}(\alpha)\!=\!\frac{1}{4\pi}\sum_{n=0}^{+\infty}\alpha^{2n}\br{
S_n C_n - S_{n+1} C_{n+1} }$, where
$S_n\!\!\!=\!\!\int_0^\pi\!\sin^{2n+1}\!\theta\ud{\theta}\!=\!2\frac{(2n)!!}{(2n+1)!!}$;  $C_n\!\!\!=\!\!\int_0^{2\pi}\cos^{2n}\!\phi\ud{\phi}\!=\!2\pi\frac{(2n-1)!!}{(2n)!!}$.
Now, $C_nS_n\!=\!\frac{4\pi}{2n+1}$ and $C_{n+1}S_{n+1}\!=\!\frac{4\pi}{2n+3}$. Hence,
$\mathcal{I}(\alpha)\!=\!\frac{1}{2\alpha}\sum_{n=0}^{\infty}\frac{2\,\alpha^{2n+1}}{2n+1}-\frac{1}{2\alpha^3}\br{-2\alpha+\sum_{n=0}^{\infty}\frac{2\alpha^{2n+1}}{2n+1}}$, where 
we have subtracted the excess term $2\alpha$ in the second series after renaming $n\!\to\! n+1$.  Both of the infinite series are 
Taylor series expansions of $\ln{{\frac{1+\alpha}{1-\alpha}}}$, therefore $$\mathcal{I}(\alpha)=\frac{1}{\alpha^2}-\frac{1-\alpha^2}{2\alpha^3}\ln{\br{\frac{1+\alpha}{1-\alpha}}}, \quad 0\leq \alpha<1.$$ Next, using the earlier expression for $\avg{\tilde{v}_r^2}$ and substituting the definition of $\alpha(r)$ in place of $\alpha$, we finally obtain \eqref{eq:mytransform}.

\section{\label{app:eqsderiv}Derivation of the equations of motion from a nonstandard Hamiltonian}
In this section we use cylindrical coordinates $(\rho,\varphi,\zeta)$.
By axial symmetry of the potential $\Psi(\rho,\zeta)$ the $J_{\varphi}$ component of the angular momentum is conserved. Thus, we may use $\varphi$ as the independent parameter for trajectories with $J_{\varphi}\ne0$.  Then the Lagrangian for a test body of mass $m$ attains the form 
$L=\frac{m}{2}\,\frac{\d{\rho}{\varphi}^2+\rho ^2+\d{\zeta}{\varphi}^2}{\d{t}{\varphi}}
-m\, {\Psi}\br{\rho ,\zeta }\d{t}{\varphi}.$ 
The dynamical variables, which we now regard as functions of $\varphi$,
are $\rho$, $\zeta$ and the time variable $t$. 
The Hamiltonian is found, as usual, by means of the Legendre transformation
$L\to
G\equiv\d{\rho}{\varphi}\pdd{\rho}{\varphi}+\d{\zeta}{\varphi}\pdd{\zeta}{\varphi}+\d{t}{\varphi}\pdd{t}{\varphi}-L=-\frac{m\,\rho ^2}{\d{t}{\varphi}}$, where the velocities must be expressed in terms of positions and momenta.
Solving  the canonical definitions of momenta
$p_\rho=\pdd{\rho}{\varphi}=m\frac{\d{\rho}{\varphi}}{\d{t}{\varphi}}$, $p_\zeta=\pdd{\zeta}{\varphi}=m\frac{\d{\zeta}{\varphi}}{\d{t}{\varphi}}$, and $p_t=\pdd{t}{\varphi}=-
m\frac{
\d{\rho}{\varphi}^2+
\rho ^2+
\d{\zeta}{\varphi}^2
}{2\d{t}{\varphi}^2} -m{\Psi}\br{\rho ,\zeta }$ for velocities,
we find that the Hamiltonian $H$ reads $$H=\pm
\rho \sqrt{-\br{p_\rho ^2+2m\,p_t +p_\zeta ^2+2m^2
\,{\Psi}\br{\rho ,\zeta }}}.$$ 
For solutions, this Hamiltonian equals minus the third component of the angular momentum (!).
Since $H$ is not an explicit function of $\varphi$, $H$ is constant for solutions.
Let denote this constant by $-J_{\varphi}$.  
Because there are two solutions for $H$ with the opposite sign and 
$H$ is constant for solutions,
it is more convenient to rewrite Hamiltonian equations into the reduced form
$\d{q}{\varphi}\!=\!-\frac{1}{2J_{\varphi}}\pb{q}{H^2}$, $\d{p}{\varphi}\!=\!-\frac{1}{2J_{\varphi}}\pb{p}{H^2}$,
where 
$\{\,,\,\}$ is the Poisson bracket on the phase space $(\rho,p_{\rho},\zeta,p_{\zeta},t,p_t)$. In explicit form these equations 
read:
$\d{\rho}{\varphi}=\frac{\rho ^2}{J_{\varphi}} p_\rho$, 
$\d{\zeta}{\varphi}=\frac{\rho ^2}{J_{\varphi}} p_\zeta$,
$\d{t}{\varphi}=\frac{\rho ^2}{J_{\varphi}}m$,
$\d{p_\rho}{\varphi}=\frac{J_{\varphi}}{\rho }-\frac{m^2\,\rho ^2}{J_{\varphi}}\,\pd{{\Psi}}{\rho }$,
$\d{p_\zeta}{\varphi}=-\frac{m^2\,\rho ^2}{J_{\varphi}}\,\pd{{\Psi}}{\zeta }$,
$\d{p_t}{\varphi}=0$.
The last equation states that $p_t$ is conserved. We denote this constant value by $-E$. Then, using the expression for $H^2$ or the definition of $p_t$ along with the Hamiltonian equations, it follows that
$E=\frac{1}{2m}\br{ {p_\rho ^2}{}+{p_\zeta ^2}{}+{J_{\varphi}^2}/\rho^2}+m\,{\Psi}(\rho ,\zeta )$.
The solution for function $t(\varphi)$ is not important for our purposes -- the time averages
along a trajectory may be expressed as averages over the angle $\varphi$, namely
 $\frac{1}{T}\int_0^T F(t) \mathrm{d}t=\frac{ \int
F \,\rho ^2\,\mathrm{d}\varphi}{\int
\rho ^2\,\mathrm{d}\varphi}$. Therefore, given constants $J_{\varphi}$ and
$E$, and having specified the initial conditions consistently with $E$ and $J_{\varphi}$, only 
four independent equations are left to be solved. On expressing $p_{\rho}$ and $p_{\zeta}$ by the physical velocities $v_{\rho}\equiv
\frac{\d{\rho}{\varphi}}{\d{t}{\varphi}}$ and $v_{\zeta}\equiv
\frac{\d{\zeta}{\varphi}}{\d{t}{\varphi}}$, the remaining first order equations for four functions $v_{\rho}(\varphi)$, $v_{\zeta}(\varphi)$, $\rho(\varphi)$ and $\zeta(\varphi)$ reduce to: 
\begin{align*}
\d{\rho}{\varphi}&=\frac{\rho ^2}{J_{\varphi}/m} v_\rho, &  \d{v_\rho}{\varphi}&=
\frac{J_{\varphi}/m}{\rho }-
\frac{\rho ^2}{J_{\varphi}/m}\,\pd{{\Psi}(\rho ,\zeta )}{\rho },\\
\d{\zeta}{\varphi}&=\frac{\rho ^2}{J_{\varphi}/m} v_\zeta, & \d{v_\zeta}{\varphi}&=-
\frac{\rho ^2}{J_{\varphi}/m}\,\pd{{\Psi}(\rho ,\zeta )}{\zeta},
\end{align*} and the energy integral attains the form  
$$E/m=\frac{v_\rho ^2}{2}+\frac{v_\zeta ^2}{2}+\frac{(J_{\varphi}/m)^2}{2\,\rho ^2}+{\Psi}(\rho ,\zeta ).$$

\bibliography{simulation}
\bibliographystyle{aa}
\end{document}